\documentclass[12pt]{article}
\usepackage{amssymb,amsmath,amsthm, amstext}
\usepackage{latexsym}
\usepackage{natbib}
\usepackage{amsbsy}
\usepackage{graphicx}
\usepackage{float}
\usepackage{caption}
\usepackage{subcaption}
\usepackage{epsfig}
\usepackage{epstopdf}  
\usepackage{color}
\usepackage[font= {small,it} ]{caption}
\usepackage{url}
\usepackage[titletoc]{appendix}
\usepackage{xcolor}  
\usepackage{hyperref}
\usepackage{appendix}

\newcommand{\B}{\boldsymbol}

\newcommand*\samethanks[1][\value{footnote}]{\footnotemark[#1]}

\usepackage{setspace}

\newenvironment{frcseries}{\fontfamily{frc}\selectfont}{}

\usepackage{JASA_manu}

\usepackage{xr}

\makeatletter
\newcommand*{\addFileDependency}[1]{
  \typeout{(#1)}
  \@addtofilelist{#1}
  \IfFileExists{#1}{}{\typeout{No file #1.}}
}
\makeatother

\graphicspath{ {Figures/} }

\date{}

\begin{document}

\title{Bayesian Spatial Binary Regression for Label Fusion in Structural Neuroimaging}
\author{}
\author{D. Andrew Brown\thanks{School of Mathematical and Statistical Sciences, Clemson University, Clemson, SC 29634, USA} \and Christopher S. McMahan\samethanks \and Russell T. Shinohara\thanks{Penn Statistics in Imaging and Visualization Center, Department of Biostatistics, Epidemiology, and Informatics, and Center for Biomedical Image Computing and Analytics, Department of Radiology, Perelman School of Medicine, University of Pennsylvania, Philadelphia, PA 19104, USA} \and Kristin A. Linn\samethanks \\[8pt] for the Alzheimer's Disease Neuroimaging Initiative\footnote{Data used in preparation of this article were obtained from the Alzheimer's Disease
Neuroimaging Initiative (ADNI) database (adni.loni.usc.edu). As such, the investigators
within the ADNI contributed to the design and implementation of ADNI and/or provided data
but did not participate in analysis or writing of this report. A complete listing of ADNI
investigators can be found at:
\url{http://adni.loni.usc.edu/wp-content/uploads/how_to_apply/ADNI_Acknowledgement_List.pdf}}}

\maketitle
\newpage

\begin{abstract}
Alzheimer's disease is a neurodegenerative condition that accelerates cognitive decline relative to normal aging. It is of critical scientific importance to gain a better understanding of early disease mechanisms in the brain to facilitate effective, targeted therapies. The volume of the hippocampus is often used in diagnosis and monitoring of the disease. Measuring this volume via neuroimaging is difficult since each hippocampus must either be manually identified or automatically delineated, a task referred to as segmentation. Automatic hippocampal segmentation often involves mapping a previously manually segmented image to a new brain image and propagating the labels to obtain an estimate of where each hippocampus is located in the new image. A more recent approach to this problem is to propagate labels from multiple manually segmented atlases and combine the results using a process known as label fusion. To date, most label fusion algorithms employ voting procedures with voting weights assigned directly or estimated via optimization. We propose using a fully Bayesian spatial regression model for label fusion that facilitates direct incorporation of covariate information while making accessible the entire posterior distribution. Our results suggest that incorporating tissue classification (e.g, gray matter) into the label fusion procedure can greatly improve segmentation when relatively homogeneous, healthy brains are used as atlases for diseased brains. The fully Bayesian approach also produces meaningful uncertainty measures about hippocampal volumes, information which can be leveraged to detect significant, scientifically meaningful differences between healthy and diseased populations, improving the potential for early detection and tracking of the disease.\\

%

\begin{keywords}
Alzheimer's disease, chromatic Gibbs sampling, conditionally autoregressive model, hippocampus segmentation, regions of interest
\end{keywords}
\end{abstract}

\newpage

\section{Introduction}
\label{sec:Introduction}


Alzheimer's is a neurodegenerative disease characterized by the deterioration of brain tissue and accelerated cognitive decline. There is no cure for Alzheimer's disease (AD). Hence, it is important to gain a better understanding of disease mechanisms in the brain and to identify biomarkers that are effective for early detection. The association between atrophy of the hippocampus and AD is well-documented \citep{Lik+etal:05, Leu+etal:10} and volumetric changes over time can be used to aid the diagnosis of early AD and to track progression \citep{Dub+etal:07}. Quantifying and tracking such changes requires delineating and measuring the hippocampus in vivo from magnetic resonance (MR) brain images of both healthy subjects and those that have been diagnosed with AD.

The manual delineation of medical images into regions of interest (ROIs) such as the hippocampus is often time-intensive and costly. Thus, even for moderate-sized research studies, obtaining manual segmentations for each image can be infeasible \citep{Igl+Sab:15}. These so-called ``gold standard'' segmentations provided by human experts, and the resulting ROI volumes, are known to exhibit intra- and inter-rater variability \citep{Yus+etal:06}. In light of these limitations, methods for automatic and semi-automatic segmentation of ROIs are critical for increasing the feasibility of imaging studies.

An {\em atlas} is an image for which a labeling exists that associates to each voxel a single label indicating the structure or region to which it belongs. Early atlas-guided segmentation techniques generally employed a single atlas to segment a single target or set of target images \citep{PhamEtAl00}. The anatomical variability of even healthy brains cannot be captured by using a single atlas. As a result, methods for using information from multiple atlases have been introduced, including best atlas selection \citep{RohlfingEtAl04} and mapping images to a single atlas of label probabilities \citep{AshFrist05}. These approaches gave way to multi-atlas segmentation (MAS). A particularly important application of MAS is segmentation and volumetry of the hippocampus \citep{Yus+etal:06, IglesiasEtAl13}. \citet{DoshiEtAl16} propose MAS via ensembles of different registration algorithms, as opposed to applying the same registration algorithm to every atlas. \citet{HuoEtAl16} propose the MaCRUISE algorithm that simultaneously performs MAS and cortical surface reconstruction. \citet{Igl+Sab:15} review MAS methods. Recently, deep learning also has been applied to accomplish brain image segmentation \citep{AkkusEtAl17, MilletariEtAl17, WachingerEtAl18}. MAS is characterized by combining information from a collection of atlases for segmenting a novel, unlabeled image through the following steps: 1) atlas generation, 2) image registration and label propagation, and 3) label fusion. Our work here is focused on the label fusion portion of MAS and assumes that multiple candidate segmentations are available in the target image space.

A popular approach to label fusion is majority voting \citep{KleinEtAl05, HeckemannEtAl06}, which takes the mode of the label distribution at each voxel after registration. An early extension to simple majority voting is weighted voting, where each atlas' `vote' is assigned a weight based on its estimated reliability. These weights may be determined by mutual information \citep{ArtaEtAl08} or iterative re-weighting based on estimated performance of each atlas \citep{LangerakEtAl10}. Local weighting was introduced to account for the fact that the quality of a given registration can vary spatially over the image. Common approaches include weighting by the voxelwise absolute image intensity differences \citep{IsgumEtAl09} or the registration Jacobian \citep{RamusEtAl10}. Joint label fusion \citep[JLF;][]{JLF13} is the current state-of-the-art MAS method. JLF weights atlases by exploiting the fact that different atlases may be correlated in terms of their label errors.

Bayesian approaches to label fusion have also been proposed in the literature. One of the first is that proposed by \cite{SabuncuEtAl10}, which posits a latent membership field to which each voxel belongs. A similar approach is taken by the STAPLE algorithm \citep{War+etal:04} in which the observed rater segmentations are treated as corrupted versions of an underlying true segmentation. \cite{AkhondiEtAl14} extend the STAPLE algorithm by associating estimated reliabilities with the membership field. To date, most Bayesian approaches to label fusion are treated as optimization problems to find the maximum a posteriori (MAP) estimator. That is, they find the mode as a point estimate but stop short of quantifying the uncertainty associated with the estimated classification. The MAP estimate can be misleading, especially when the posterior distribution is widely dispersed or strongly skewed \citep{FoxNicholls01}. For example, \citet{GreenlawEtAl17} illustrate an imaging genetics study in which a particular single nucleotide polymorphism was identified as being associated with all imaging phenotypes via MAP point estimation, but was subsequently shown to be not statistically significant after accounting for uncertainty via credible intervals. \color{black} It thus is desirable to access the entire posterior distribution whenever possible, not only to estimate a posterior probability map for segmentation, but to have available any desired summary measures along with appropriate measures of variability.

Modern label fusion procedures use atlas-target image agreement as auxiliary information to guide segmentation. However, additional information such as the tissue class of each voxel can be useful, since it is known that the hippocampus consists only of gray matter. In this work we construct a fully Bayesian, latent variable regression model that incorporates covariate information via a generative model for hippocampus segmentation. This is distinct from voting-based label fusion procedures that posit discriminative models and thus do not incorporate relevant information that is independent of any registered atlas. \color{black}  We accommodate the variability of image registration quality both within and between rater images by using spatially-varying sensitivity and specificity processes, thereby facilitating smooth, local weighting of each image as an inherent part of the model. We discuss prior elicitation and implementation of the model via Markov chain Monte Carlo \citep[MCMC;][]{GelfandSmith90}. We find that including tissue class segmentations in addition to image agreement can be particularly useful for segmenting diseased brains. This is important since segmenting diseased cases is more challenging than standard segmentation problems due to the heterogeneity of the target images.

We motivate and present our proposed approach in Section \ref{sec:Methods}, along with practical considerations such as prior elicitation and implementation. In Section \ref{sec:Simulation} we consider simulated examples with corrupted segmentations to compare its performance to that of simple and weighted majority voting against a known truth, as well as to illustrate the potential value of auxiliary information. In Section \ref{sec:Application} we present our clinical MRI data obtained from both healthy subjects and those that have been diagnosed with Alzheimer's disease. We demonstrate that existing label fusion procedures can have difficulty segmenting diseased cases and contrast them against our approach, which is shown to be useful for estimating plausible volumes of a diseased hippocampus, including measures of uncertainty about each volume. Section \ref{sec:Discussion} concludes with a discussion and thoughts about future research directions.  All of the code for reproducing the results in Sections \ref{sec:Simulation} and \ref{sec:Application} may be found as supplementary material.\color{black}

\section{Methods}
\label{sec:Methods}

\subsection{Model}
\label{subsec:Model}

\noindent
Suppose we are interested in segmenting a new brain image, hereinafter referred to as the target image. We have available $R$ previously manually segmented images to use as atlases after aligning the labeled brains to the target. This alignment is done via {\em image registration}, the process of transforming one image to the space of another via an affine or nonlinear map. The registration problem is itself an ongoing area of research in the imaging literature, but state of the art tools include the FLIRT and FNIRT functions in FSL \citep{JenkSmith01, JenkinsonEtAl02, GreveFischl09, JenkinsonEtAl12}, DRAMMS \citep{OuEtAl11}, LDDMM \citep[e.g.,][]{ZhongEtAl10}, and SyN registration \citep{TustAvants13}. In what follows, we assume that a single algorithm is used to register each atlas to the target, but our proposed framework can easily account for different registration algorithms. In either case, the sources of uncertainty leading to corruption of the atlas labels are the radiologists' imperfect manual segmentations, the set of brains that have been manually segmented, none of which is identical to the target brain, and the registration algorithm itself, which cannot produce a perfect alignment between images.

After registering each atlas to the target, we have $R$ images indexed by $v= 1, \ldots, V$, where $V$ is the number of voxels. Corresponding to atlas $r \in \{1, \ldots, R\}$ is a set of labels $\{Y_{1r}, \ldots, Y_{Vr}\} \in \{0, 1\}^V$. The atlases will disagree with each other after registration and will each be a corrupted version of the underlying truth. In addition to the observed labels $Y_{vr}$, let $T_v \in \{0,1\}, ~v= 1, \ldots, V$, denote the ``true" (unobservable) status of the voxel $v$, so that $T_v=1$ indicates that voxel $v$ is a part of the structure of interest, and $T_{v}=0$ otherwise. 

We associate to each atlas $r$ a {\em sensitivity} $\xi$, an ability to correctly detect when a voxel is truly part of the structure of interest. Similarly, we have also for each atlas a {\em specificity} $\psi$, an ability to correctly determine when a voxel is truly outside of the ROI. Since the quality of any registration can vary throughout an image, the best atlas to use depends on location \citep{ArtaEtAl09}. Thus, we allow the sensitivity and specificity of each atlas to be a function not only of the specific atlas, but also the voxel. We suppose that $P(Y_{vr} = 1 \mid T_{v} = 1)  =  \xi(v,r)$ and $P(Y_{vr} = 0 \mid T_{v} = 0) =  \psi(v,r)$, for atlases $r= 1, \ldots, R$ and voxels $v= 1, \ldots, V$. Given the true voxel statuses and the sensitivity and specificity of each atlas, we assume that observed labels are conditionally independent,
\begin{equation}\label{eqn:datDist}
    Y_{vr} \mid T_v, ~\xi(v,r), ~\psi(v,r)  \stackrel{ind.}{\sim}  \text{Bern}(p^\ast(v,r)), ~~v= 1, \ldots, V; ~r= 1, \ldots R,
\end{equation}
where $p^\ast(v,r) = \xi(v,r)^{T_{v}}[1-\psi(v,r)]^{1-T_{v}}$. For example, if $Y_{vr_0}$ and $Y_{v^\prime r_0}$ come from the same atlas $r_0$, we assume the dependence between them is only due to the dependence between $\xi(v,r_0)$ and $\xi(v^\prime, r_0)$ and between $\psi(v,r_0)$ and $\psi(v^\prime, r_0)$.

We propose a framework for incorporating spatial smoothness into the sensitivity and specificity processes, as well as additional covariate information that can be informative with respect to the true voxel status $T_v$. The spatial process and covariate models themselves depend on unobservable and unknown parameters. Hence we take a hierarchical Bayesian approach and assign them prior distributions as well. The prior distributions and concomitant hyper-parameters are described in this subsection.

It is important to impose smoothness on the reliabilities to mitigate the effect of image noise \citep{ArtaEtAl09}. We model the sensitivities and specificities as
\begin{eqnarray}\label{eqn:Sens}
    \begin{aligned}
        \xi(v,r) & = \Phi(\B{x}_{vr}^\top\B{\beta}_r + \phi_{vr});
        ~~\psi(v,r) =  \Phi(\B{z}_{vr}^\top\B{\gamma}_r + \eta_{vr}),
    \end{aligned}
\end{eqnarray}
where $\Phi(\cdot)$ is the standard Gaussian CDF, $\B{x}_{vr} \in \mathbb{R}^\ell$ and $\B{z}_{vr} \in \mathbb{R}^k$ are known vectors of covariates with coefficients $\B{\beta}_r$ and $\B{\gamma}_r$, respectively, and $\phi_{vr}$ and $\eta_{vr}$ are elements of mean-zero Gaussian Markov random fields \citep[GMRFs;][]{RueHeld05}. Our experience is that it is often sufficient to simply set $\B{x} = \B{z} = \B{0}$ to let the spatial processes detect the trends {\em a posteriori}. Observe that we assume that $\xi(v,r)$ and $\psi(v,r)$ are independent of each other, since they are defined by two mutually exclusive outcomes, $T_v = 1$ and $T_v = 0$.

We assume $\boldsymbol{\phi}_{r}=(\phi_{1r},...,\phi_{Vr})^\top \in \mathbb{R}^V$ and $\boldsymbol{\eta}_{r}=(\eta_{1r},...,\eta_{Vr})^\top \in \mathbb{R}^V$, $r=1,...,R$, each independently follow conditionally autoregressive models \citep[CAR;][]{Besag74}, given by
\begin{eqnarray}\label{eqn:CARmods}
    \begin{aligned}
        \boldsymbol{\phi}_{r} & \stackrel{ind.}{\sim}  N_{V}\left\{\B{0}, \tau_{\B{\phi}_r}^{-1} (\B{D}-\rho_{\B{\phi}_r} \B{W})^{-1} \right\};
        ~~\boldsymbol{\eta}_{r}  \stackrel{ind.}{\sim}  N_{V}\left\{\B{0}, \tau_{\B{\eta}_r}^{-1} (\B{D}-\rho_{\B{\eta}_r} \B{W})^{-1} \right\}.
    \end{aligned}
\end{eqnarray}
Here, $\B{W} = \{w_{ij}\}_{i,j=1}^V \in \mathbb{R}^{V \times V}$ is a neighborhood matrix such that $w_{v,v^\prime} = I(v \sim v^\prime)$, where $v \sim v^\prime$ if and only if voxels $v$ and $v^\prime$ share an edge or a corner, $I(\cdot)$ is the indicator function, and $\B{D} = \text{diag}\left(\sum_{j=1}^Vw_{ij}, ~i= 1, \ldots, V\right) \in \mathbb{R}^{V\times V}$. This yields a nonstationary process due to the edge effects; i.e., the voxels on the edges have different numbers of neighbors than the interior voxels, leading to different marginal variances. \cite{BesagKoop95} observe that when images are of large dimension and the regions of interest are in the interior, edge effects are negligible with respect to inferences, so one can safely ignore them.

When $T_{v} = 0$ for all $v$, $\xi(v,r)$ does not appear in the likelihood determined by \eqref{eqn:datDist} and hence $\B{\phi}_r$ is not updated in the posterior. Since $T_{1} = \cdots = T_{V} = 0$ with positive probability, the prior distribution on $\B{\phi}_r$ must be proper to avoid an improper posterior, and likewise for $\B{\eta}_r$. Thus, we include the ``propriety parameters" $\rho_{\B{\phi}_r}$ and $\rho_{\B{\eta}_r}$ in \eqref{eqn:CARmods} to force the precision matrices to be nonsingular, as suggested by \citet{BanerjeeEtAl15}. A sufficient condition is $|\rho_{\B{\phi}_r}|  < 1$, in which case $(\B{D}-\rho_{\B{\phi}_r} \B{W})$ is diagonally dominant and hence positive definite. 



It is likely that we will have available auxiliary information concerning the true location of the structure of interest. Toward this end, we suppose that,
\begin{equation}\label{eqn:TruStats}
   P(T_{v} = 1 \mid \boldsymbol{\delta}) = g^{-1}(\B{c}_{v}^\top\boldsymbol{\delta}), ~~v= 1, \ldots, V,
\end{equation}
where $\B{c}_{v} \in \mathbb{R}^J$ is a vector of covariates pertaining to voxel $v$, $\B{\delta}$ is the corresponding vector of regression coefficients, and $g: (0,1) \rightarrow \mathbb{R}$ is a one-to-one and differentiable link function. We assume any dependence between inclusion indicators is completely explained by the covariate information; i.e., $T_v \perp T_{v^\prime} \mid \B{\delta}, ~v \neq v^\prime$. In Sections \ref{sec:Simulation} and \ref{sec:Application}, we demonstrate the use of signed distance label maps and either normalized image intensity or tissue class segmentations as covariate information for $\B{c}_v$. In Subsection \ref{sub:priors}, we discuss prior elicitation for $\B{\delta}$.

\begin{figure}[tb]
    \centering
    \includegraphics[scale=.45, clip= TRUE, trim= 0in 1.75in 0in 2.45in]{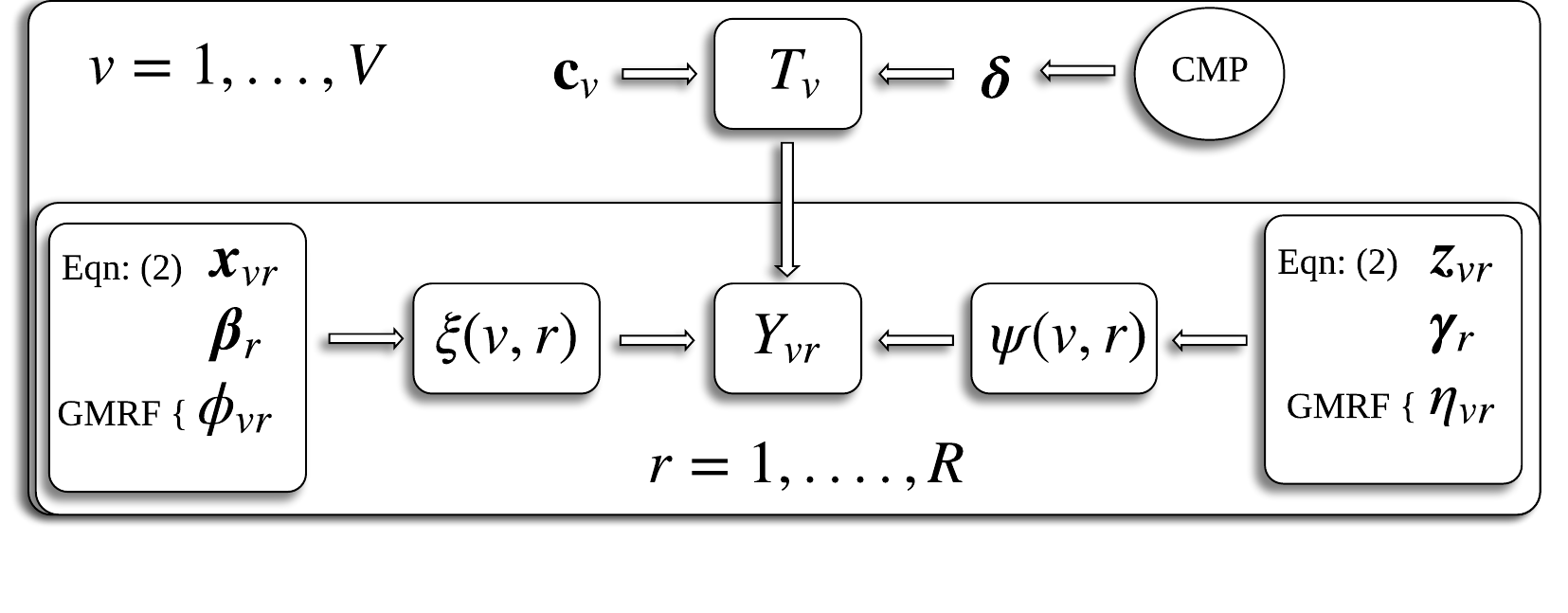}
    \caption{Graphical depiction of the proposed Bayesian label fusion model for segmenting brain images consisting of $V$ voxels using $R$ available atlases.}
    \label{fig:modelGraph}
\end{figure}

The model is completed with prior distributions on the regression coefficients and precision parameters in \eqref{eqn:Sens}, \eqref{eqn:CARmods}, and \eqref{eqn:TruStats}. We take conventional Gaussian and Gamma priors for these parameters, since they are sufficiently flexible with appropriate specification of the corresponding hyperparameters. Thus, we assume
\begin{eqnarray}\label{eqn:hypPriors}
    \begin{aligned}
        \B{\beta}_r  \stackrel{iid}{\sim} N_{\ell}\left(\B{0},\B{\Sigma}_{\B{\beta}}\right); ~~\B{\gamma}_r &\stackrel{iid}{\sim} N_{k}\left(\B{0},\B{\Sigma}_{\B{\gamma}}\right), ~r= 1, \ldots, R \\
        \B{\delta}  &{\sim} N_{J}\left(\B{0},\B{\Sigma}_{\B{\delta}}\right),\\
        \tau_{\B{\phi}_r} \stackrel{iid}{\sim} Ga\left(a_{\B{\phi}}, b_{\B{\phi}}\right); ~~\tau_{\B{\eta}_r} &\stackrel{iid}{\sim} Ga\left(a_{\B{\eta}}, b_{\B{\eta}}\right), ~r= 1, \ldots, R.
    \end{aligned}
\end{eqnarray}
To aid with exposition, Figure \ref{fig:modelGraph} provides a graphical depiction of our proposed model. Let $\B{\omega}$ denote the collection of all model parameters except the inclusion indicators $\B{T} = (T_1, \ldots, T_V)^\top \in \mathbb{R}^V$. Under the model determined by equations \eqref{eqn:datDist} - \eqref{eqn:hypPriors}, the joint posterior density of the parameters, conditional on the observed data $\B{Y} = (Y_{11}, \ldots, Y_{VR})^\top \in \mathbb{R}^{VR}$, is
\begin{eqnarray*}
\pi(\B{T}, \B{\omega} \mid \B{Y}) & \propto & \prod_{v=1}^{V} \prod_{r=1}^{R} \Phi(\phi_{vr})^{Y_{vr}T_v}    \{1-\Phi(\phi_{vr})\}^{(1-Y_{vr})T_v} \Phi(\eta_{vr})^{(1-Y_{vr})(1-T_v)} \{1-\Phi(\eta_{vr})\}^{Y_{vr}(1-T_v)} \\
& & \times \prod_{v=1}^{V} \{g^{-1}(\B{c}_{v}^\top\boldsymbol{\delta})\}^{T_v}\{1-g^{-1}(\B{c}_{v}^\top\boldsymbol{\delta})\}^{1-T_v}N_J(\B{\delta} | \B{0}, \B{\Sigma}_{\B{\delta}})\\
& & \times \prod_{r=1}^{R} N_V(\B{\phi}_r |  \B{X}_r \B{\beta}_r, ~\tau_{\B{\phi}_r}(\B{D}-\rho_{\B{\phi}_r}\B{W})^{-1})N_V(\B{\eta}_r |  \B{Z}_r \B{\gamma}_r, ~\tau_{\B{\eta}_r}(\B{D}-\rho_{\B{\eta}_r}\B{W})^{-1}) \\
& & \times \prod_{r=1}^{R}N_{\ell}(\B{\beta}_r | \B{0}, \B{\Sigma}_{\B{\beta}})
                          N_k(\B{\gamma}_r | \B{0}, \B{\Sigma}_{\B{\gamma}})Ga(\tau_{\B{\eta}_r}|a_{\B{\eta}_r}, b_{\B{\eta}_r})Ga(\tau_{\B{\phi}_r}|a_{\B{\phi}_r}, b_{\B{\phi}_r}).     
\end{eqnarray*}

We remark that our proposed label fusion approach, particularly its ability to incorporate atlas-independent information about structure location, is considerably different from popular voting-based approaches. Each of the additional approaches we consider in Section \ref{sec:Application} estimates the probability of inclusion of voxel $v$ with $\hat{P}(T_v = 1 | \mathbf{Y}) \propto \sum_{r=1}^R w_r(v)Y_{vr},$ where $w_r(v)$ is a weight assigned to the rater $r$ classification at voxel $v$. Simple majority voting gives each atlas equal weight throughout the image ($w_r(v) \equiv 1$), globally-weighted majority voting takes $w_r(v) \equiv w_r$, and both locally-weighted majority voting and JLF produce weights that depend on both the rater ($r$) and voxel ($v$). What the weighted approaches have in common is that weights are determined via some measure of image similarity that serves as a proxy for image registration quality and thus how ``trustworthy" a particular atlas is. Put another way, each of the voting procedures and JLF are {\em discriminative} models for classification in that they aim to estimate $P(T_v | \B{Y})$ directly without accounting for any underlying process in $P(Y_{vr} | T_v)$. By contrast, our proposed approach is a {\em generative} approach that includes both pieces by classifying based on a model of the form $P(Y_{vr}, T_v) = P(Y_{vr} | T_v)\pi(T_v)$. The gray matter information is included as part of the prior information contained in $\pi(T_v)$ via a logit regression. A prior $\pi(T_v)$ is not part of any of the other procedures, and so there is no place in which we can directly incorporate gray matter or other potentially informative information derived from the target image without substantially expanding their assumed models. Both discriminative and generative classification models have their strengths and weaknesses \citep{NgJordan02}. In Section \ref{sec:Application}, we use an independent gray matter segmentation to guide segmentation, but other information could be used in different settings.\color{black}

\subsection{Prior Specification}\label{sub:priors}
To specify the hyperparameters determining the prior distributions for $\B{\beta}_r, \B{\gamma}_r$, $\B{\phi}_r$, and $\B{\eta}_r$ appearing in \eqref{eqn:Sens}, we recognize that $\Phi(4) - \Phi(-4) \approx 1$, so that the effect of the mean function on the likely values of the reliability parameters is almost certainly between $-4$ and $4$. For instance, considering the sensitivity, {\em a priori} we suppose that for any $v$ and any $r$, $P(|\B{x}_{vr}^\top\B{\beta}_r + \phi_{vr}| \leq 4) \approx 1$ for all $\B{x}_{vr}$. With a pure spatial process ($\B{x}_{vr} = \B{0}$, for all $v,r$), this requirement becomes  $P(|\phi_{vr}| \leq 4) \approx 1$. To use this to induce a prior on the spatial effect $\phi_{vr}$, we use the fact that $ Var(\phi_{vr}) \approx (0.7^2w_v.\tau_{\B{\phi}_r})^{-1}$, where $w_v. = \sum_{k=1}^Vw_{vk}$ \citep{BernardinelliEtAl95, EberlyCarlin00}. With eight neighbors at a voxel $v$, for instance, the desired constraint can be solved to yield $\tau_{\B{\phi}_r} \approx 0.5$. Hence we take $a_\phi = 1$ and $b_\phi = 2$ in \eqref{eqn:hypPriors}. The argument is the same for specifying the hyperparameters on $\B{\gamma}_r$ and $\tau_{\B{\eta}_r}$.

With prior knowledge concerning the underlying structure of interest, it is possible to induce a prior on $\B{\delta}$ through a so-called conditional mean prior \citep[CMP;][]{bedrick1996new}. Under this approach, we partition the $J$-dimensional covariate space into $J$ regions and choose hypothetical covariate vectors $\widetilde{\B{c}}_1, \widetilde{\B{c}}_2, \ldots, \widetilde{\B{c}}_J$ so that $\widetilde{\B{C}} = (\widetilde{\B{c}}_1 ~\widetilde{\B{c}}_2 ~\ldots \widetilde{\B{c}}_J)^\top \in \mathbb{R}^{J \times J}$ is nonsingular.  A prior is put on the mean response at these covariate values, $E[(\widetilde{T}_1, \ldots, \widetilde{T}_J)^\top \mid \B{\delta}] = \B{G}^\prime(\widetilde{\B{C}}\B{\delta}) ~\in \mathbb{R}^J$, where $\B{G}^\prime(\cdot)$ applies $g^{-1}$ componentwise. While the regression coefficients may be difficult to interpret, we can meaningfully assign a distribution to $\widetilde{p}_j = g^{-1}(\widetilde{\B{c}}_j^\top\B{\delta}) = P(\widetilde{T}_j = 1 \mid \B{\delta} )$ using a Beta distribution with shape parameters chosen to reflect the prior knowledge at the covariate values. \citet{bedrick1996new} show that if $\widetilde{p}_j \stackrel{indep.}{\sim} \textrm{Beta}(a_{\widetilde{p}_j},b_{\widetilde{p}_j}), ~j= 1, \ldots, J$, then the induced prior is
$\pi(\B{\delta}) \propto \prod_{j=1}^J \{g^{-1}(\widetilde{\mathbf{c}}_j^\top\B{\delta})\}^{a_{\widetilde{p}_j}-1}
\{1-g^{-1}(\widetilde{\mathbf{c}}_j^\top\B{\delta})\}^{b_{\widetilde{p}_j}-1} \dot{g}^{-1}(\widetilde{\mathbf{c}}_j^\top\B{\delta})$.



Posterior inference is facilitated via MCMC, taking advantage of data augmentation \citep{AlbertChib93} and so-called chromatic sampling \citep{BrownMc17} for updating the spatial fields. A more detailed discussion may be found in the Supplementary Material.

\section{Numerical Studies}\label{sec:Simulation}
To demonstrate the utility of incorporating auxiliary information into the label fusion procedure, we consider two controlled scenarios in which we can compare results against a known truth. First, we consider the ideal case in which each of the atlas segmentations is reliable and closely corresponds to the true structure. Covariate information is not critical in this situation, as any of the considered procedures perform well. To illustrate the benefit of including concomitant information to guide segmentation, we study in the second subsection a challenging situation in which some of the segmentations are unreliable but nevertheless tend to agree on the wrong location. This situation is similar to what can occur when healthy brain images are used as atlases for segmenting diseased brains; i.e., they might all systematically oversegment the hippocampus since healthy hippocampi tend to be larger than those that are diseased.

\subsection{The Case of Quality Atlases}
\begin{figure}[tb]
	\centering
	\includegraphics[scale= 0.35]{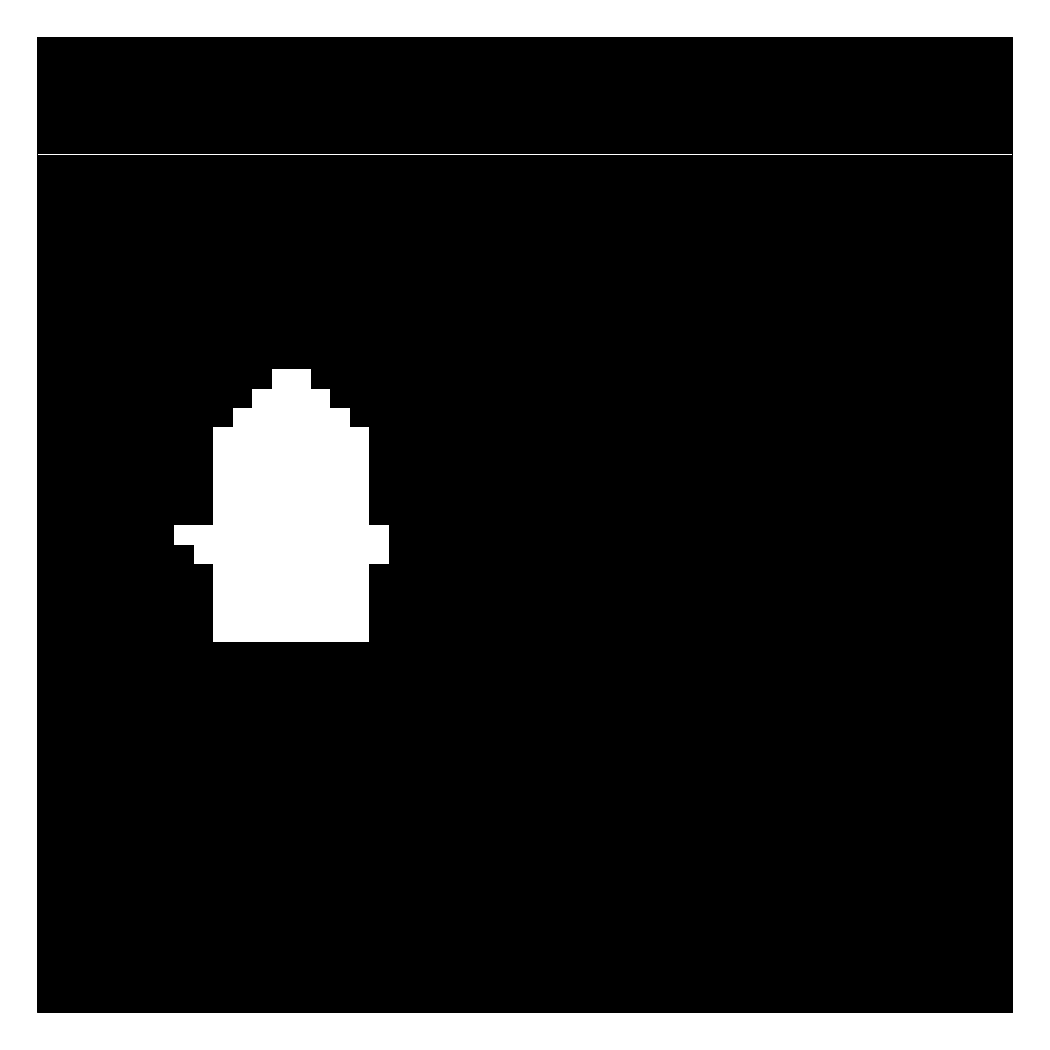}
	\includegraphics[scale= 0.35]{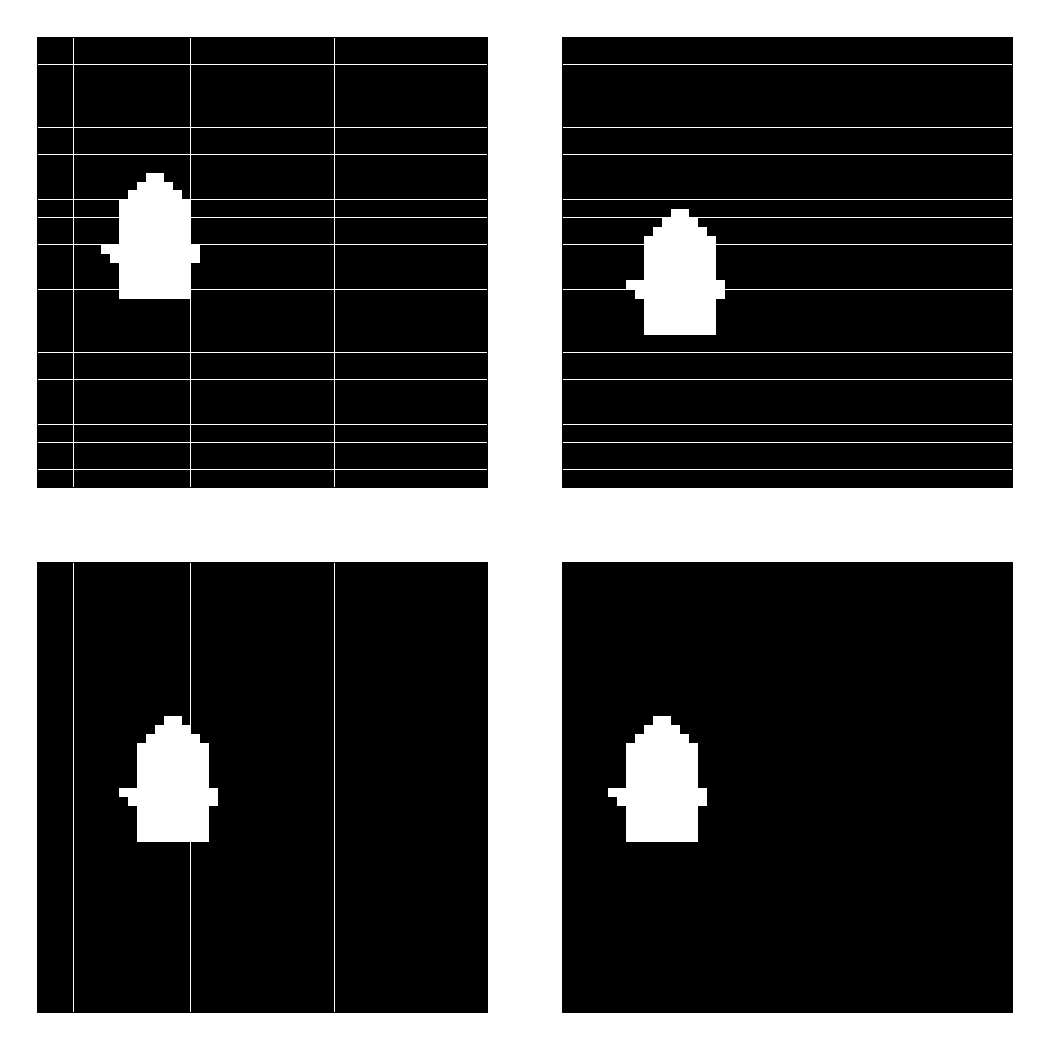}
	\caption{True structure (left) and atlases (right) in the simulated example with quality atlases. Notice that each atlas segmentation is slightly offset from the truth in a different direction.}\label{fig:GoodSegs}
\end{figure}

Here we are interested in recovering a target structure in a two-dimensional image using $R = 4$ atlases. We simulate a 40 $\times$ 40 grid, displayed in Figure \ref{fig:GoodSegs}. Suppose that each of the atlas segmentations perform well, each being only slightly offset from the truth. The atlases are also displayed in Figure \ref{fig:GoodSegs}.

To compare results of our procedure with and without covariates, we find the {\em signed distance label (SDL) transforms} for each atlas \citep{IglesiasEtAl12}. An SDL transform for a binary image assigns to each pixel a number corresponding to its distance from the nearest edge in an image, where an edge is indicated by a zero adjacent to a one. The sign of the distance corresponds to whether or not the pixel is inside or outside an identified structure; e.g., a pixel has negative distance if it is inside the structure, positive distance if it is outside, and zero if it is on the boundary. We consider the sum of the signed distance label transform maps as a possible covariate. In this case, the dimension of the predictor space with and without the covariate included is $J = 2$ and $J = 1$, respectively. In the former case, our CMP prior is induced by supposing a priori that a voxel with a large negative signed distance label has probability 0.9 of being in the structure, and probability 0.10 if its signed distance label is large positive. With no predictors at all, we induce the prior on $\B{\delta}$ by supposing there is a 50\% chance that a borderline voxel (signed distance label = 0) is truly part of the structure. Further discussion of prior elicitation for the numerical experiments, as well as MCMC implementation, is in the online Supplementary Material.

\begin{figure}[tb]
	\centering
	\includegraphics[scale= 0.5]{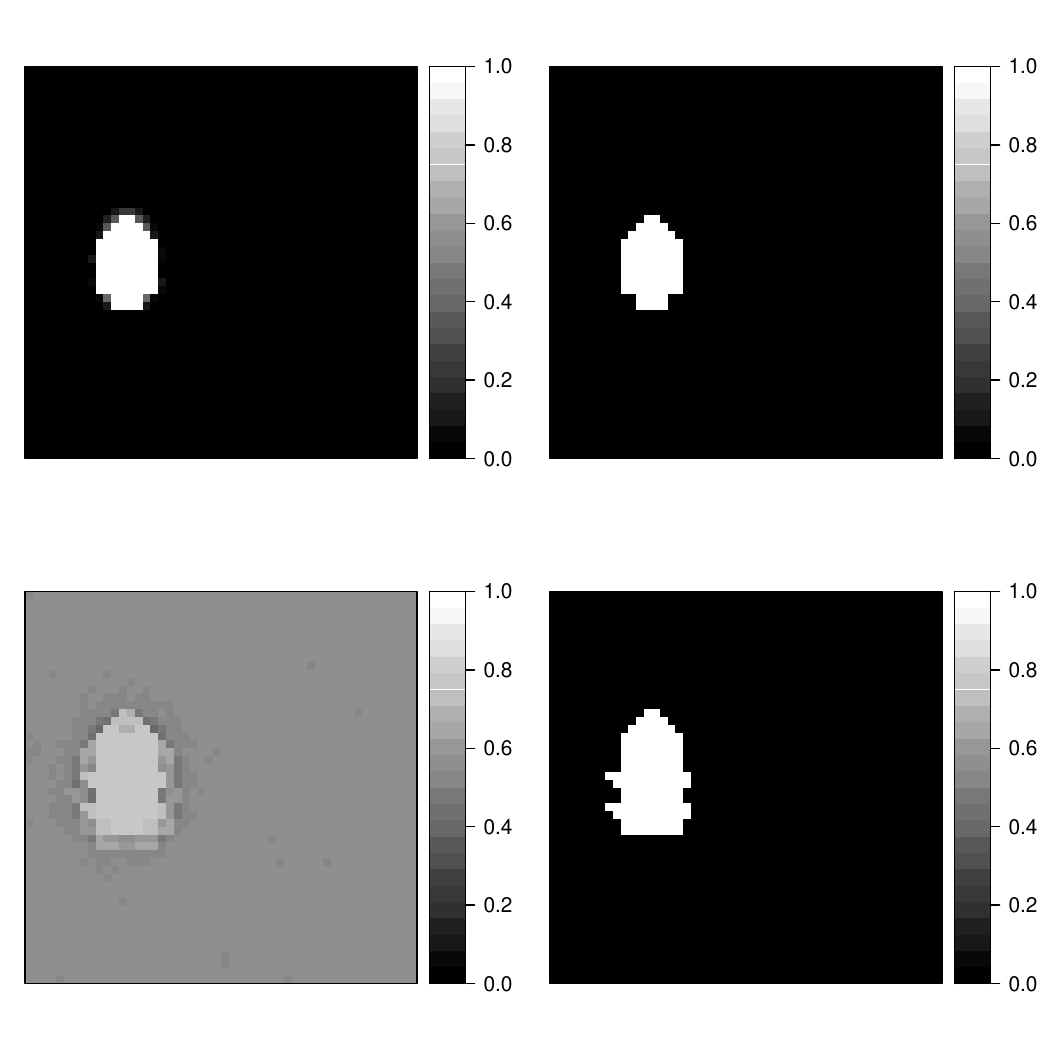}
	\caption{Bayesian label fusion results using quality atlases. The top row shows the posterior probability map (left) and the thresholded segmentation (right) using signed distance labels as a covariate. The bottom row shows the same for the intercept-only model.}\label{fig:GoodSegsBLFResults}
\end{figure}

Figure \ref{fig:GoodSegsBLFResults} displays the posterior probability maps obtained from including the signed distance labels as covariates as well as the results from using an intercept-only model. We observe a more clearly-defined region of high and low probability when the covariate information is included, whereas using no covariate information results in a less clearly defined region. The latter is a byproduct of the fact than an intercept-only  Bayesian probit regression is equivalent to allowing every pixel equal prior probability of inclusion in the structure, so that the atlases alone determine the likely region(s) of interest. 

\subsection{Using Auxiliary Information}\label{ssec:auxInfo}
It may occur in practice that multiple reference atlases are subject to the same deleterious effect, resulting in multiple atlases closely agreeing on an incorrect segmentation. If we have one atlas that registered well and provides a good segmentation, it could still be outweighed by multiple poor atlases. Thus any voting procedure would produce a poor segmentation.

\begin{figure}[tb]
    \centering
    \includegraphics[scale= 0.225]{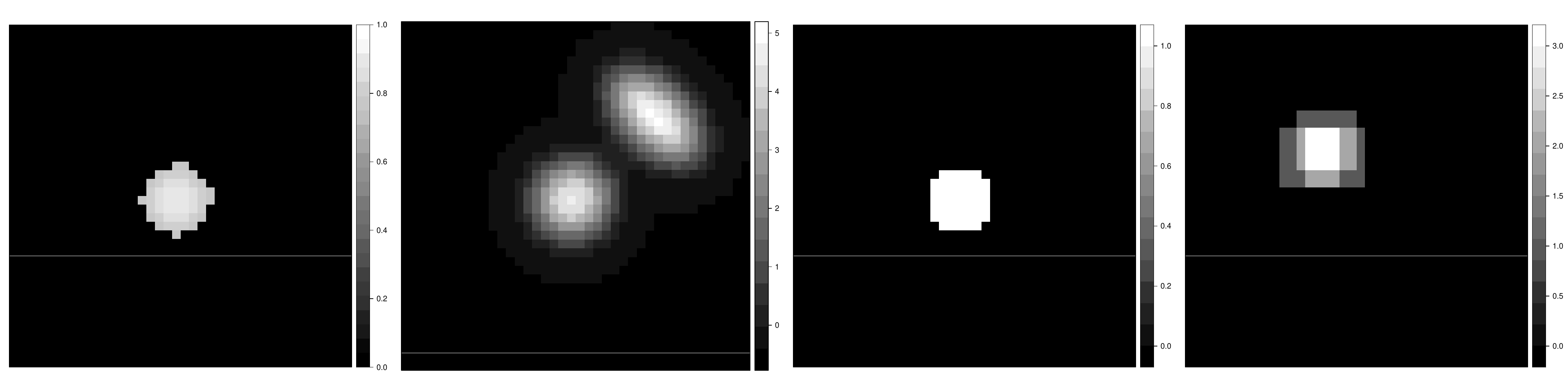}
    \caption{True structure of interest (far left), along with the associated normalized image intensities (middle left), one quality segmentation (middle right) and three poor atlas segmentations (far right) for the simulated label fusion example. The three poor segmentations are superimposed to indicate their agreement.}
    \label{fig:pvTruth}
\end{figure}

We again simulate a structure on a $40 \times 40$ grid, displayed in Figure \ref{fig:pvTruth}. Also displayed in the figure is one reliable candidate segmentation and three poor segmentations, the latter of which closely agree. We also simulate image intensities as might be obtained from an additional imaging modality. Voxelwise image intensity similarity is often used to quantify rater reliability throughout an image \citep{IsgumEtAl09}. As such, we simulate image intensity similarities for each atlas, displayed in Supplementary Figure 1. To make the assessment more realistic, we slightly offset the intensity differences from the areas of poor segmentations.

We again find the SDL transform for each atlas \citep{IglesiasEtAl12}. Here, we take a weighted sum over the atlas-specific SDL maps, where the weights are inversely proportional to the square of the intensity difference between the rater image and the target; i.e., $c_{v1} = \sum_{r=1}^4 |i_r(v) - i_t(v)|^{-2}d_r(v)/(\sum_{r = 1}^4 |i_r(v) - i_t(v)|^{-2})$, where $i_r(v)$ and $i_t(v)$ are the intensity values for rater image $r$ and the target image at pixel $v$, respectively, and $d_r(v)$ is the SDL assigned to pixel $v$ from rater image $r$. We use the pixel-level intensities as additional covariates, $c_{v2} = i_t(v)$. It is important to include a distance-intensity interaction, $c_{v3} := c_{v1}c_{v2}$, since high intensity voxels far away from the object of interest should not be included, but the information can be helpful when we believe a voxel is close to the object.

We use CMP to induce a prior on the $\B{\delta}$ coefficients in \eqref{eqn:TruStats}. The dimension of the predictor space is $J = 4$, so we augment the model with pseudo-observations under covariate values corresponding to (i) small distance, high intensity; (ii) large distance, low intensity; (iii) mid-distance and average intensity; and (iv) large distance but high intensity. We impose prior knowledge that (i) is very likely to be included in the object, (ii) is very unlikely to be included, (iii) is a borderline case, and (iv) is not very likely to be included. See the Supplementary Material for more details.

For posterior inference, a single Monte Carlo Markov chain is run for 100,000 iterations, thinning to every $25^{\text{th}}$ iterate to save memory and reduce autocorrelation. The first half of the chain is discarded as a burn-in period. We use trace plots of the $\B{\delta}$ coefficients along with Geweke statistics \citep{Geweke92} and lag 1 autocorrelations as convergence diagnostics. The diagnostics, displayed in Supplementary Figure 2, provide evidence of convergence.

\begin{figure}[tb]
    \centering
    \includegraphics[scale= 0.2, clip= TRUE]{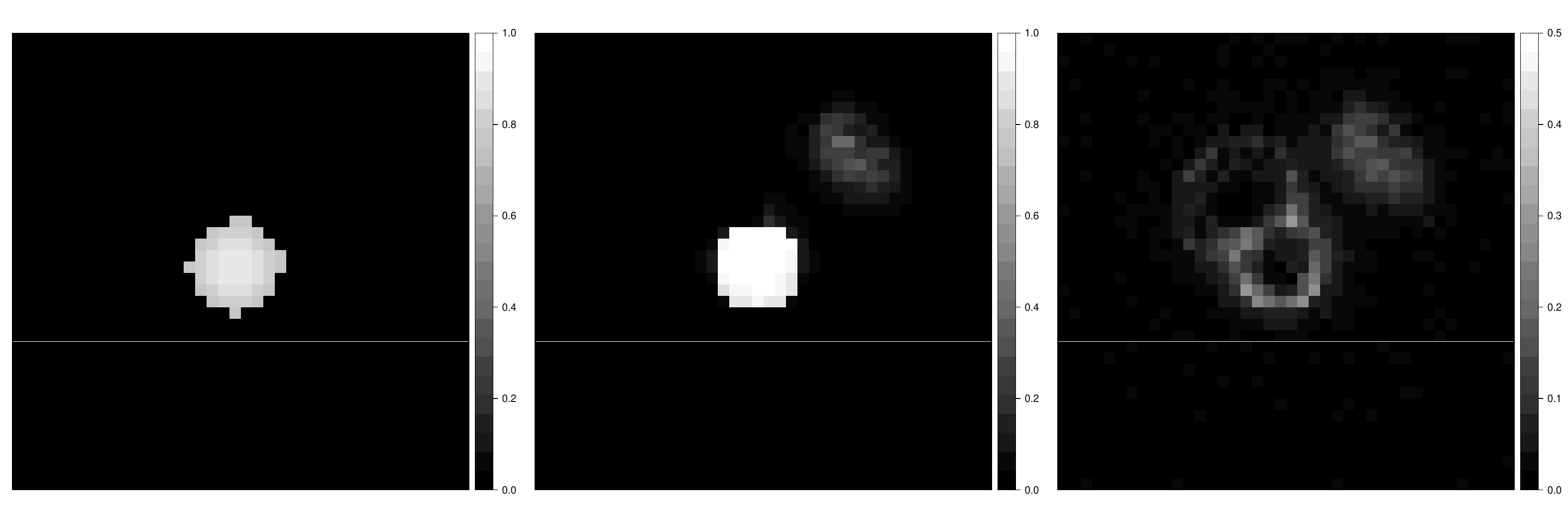}
    \caption{Target structure (left), posterior mean estimate (center) and approximate standard deviations (right) of the target structure in the simulated label fusion example. (The color scale in the right plot is different from the other two for better definition.)}
    \label{fig:RBPostMean}
\end{figure}

Figure \ref{fig:RBPostMean} displays the posterior mean estimate of the target along with pixel-wise standard deviations. To reduce the Monte Carlo variance, we use ``Rao-Blackwellized" estimators, $\hat{P}(T_v = 1 \mid \B{Y}) = N^{-1}\sum_{k=1}^N\hat{P}(T_v = 1 \mid \B{\omega}^{(k)}, \B{Y})$, where $\B{\omega}^{(k)}$ denotes the $k^{\text{th}}$ MCMC iterate of all the parameters except $T_1, \ldots, T_V$. Close agreement between the target and the estimate is evident. As expected, the uncertainty about the structure is largest near the edges of the observed segmentations, as well as in the area of higher intensity. The effect of high intensity is strongly attenuated when it is thought to be well away from the target object. Posterior inference also produces estimates of the reliabilities of the rater atlases. Supplementary Figure 3 displays the estimated sensitivity and specificity fields for each of the four atlas segmentations. We see that all atlases have high specificity in regions well away from the structure. Near the object, however, we observe high sensitivity of the good atlas with lower values for the poor ones. Conversely, the tendencies for the bad atlases to produce false positives are seen in the areas of low specificity.

\begin{figure}[tb]
    \centering
    \includegraphics[scale= 0.35, clip= true]{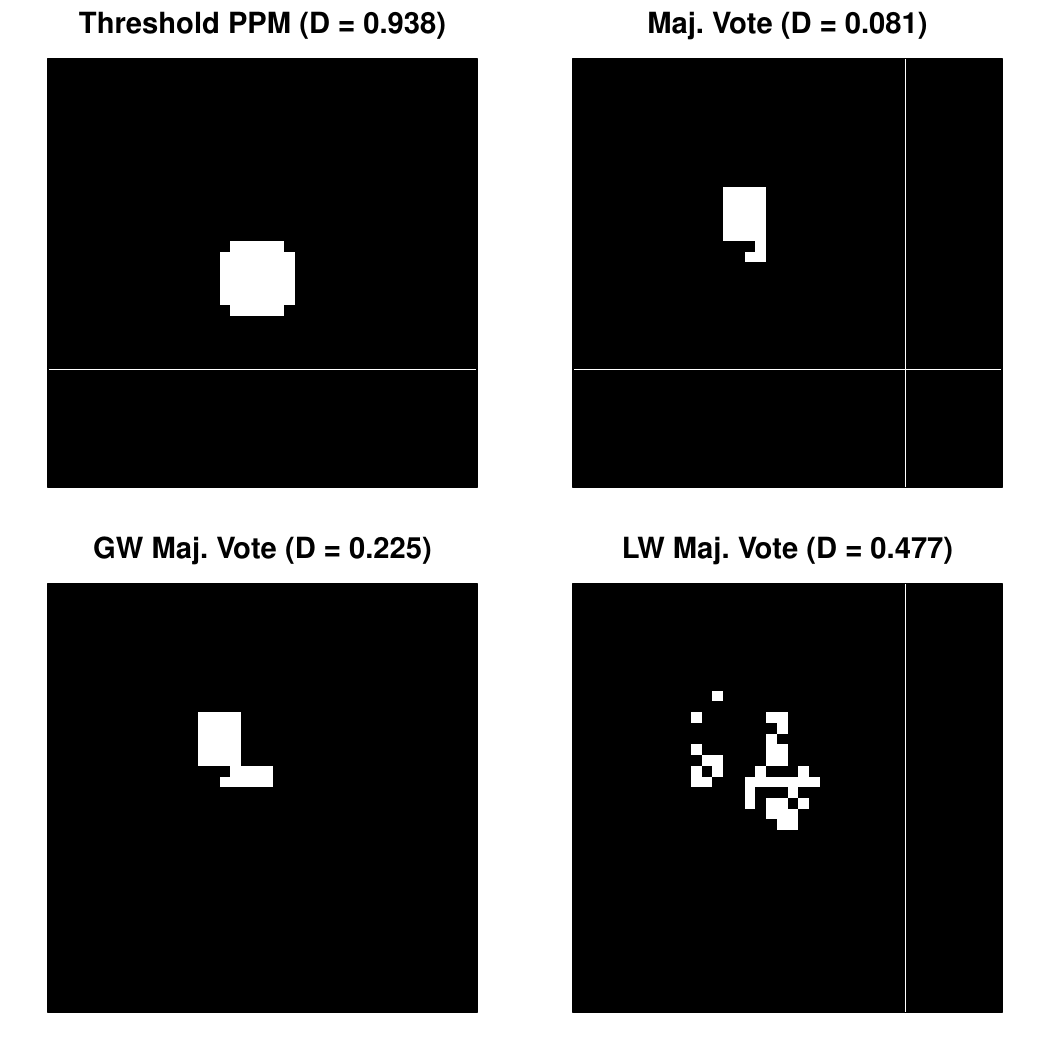}
    \includegraphics[scale= 0.35, clip= true]{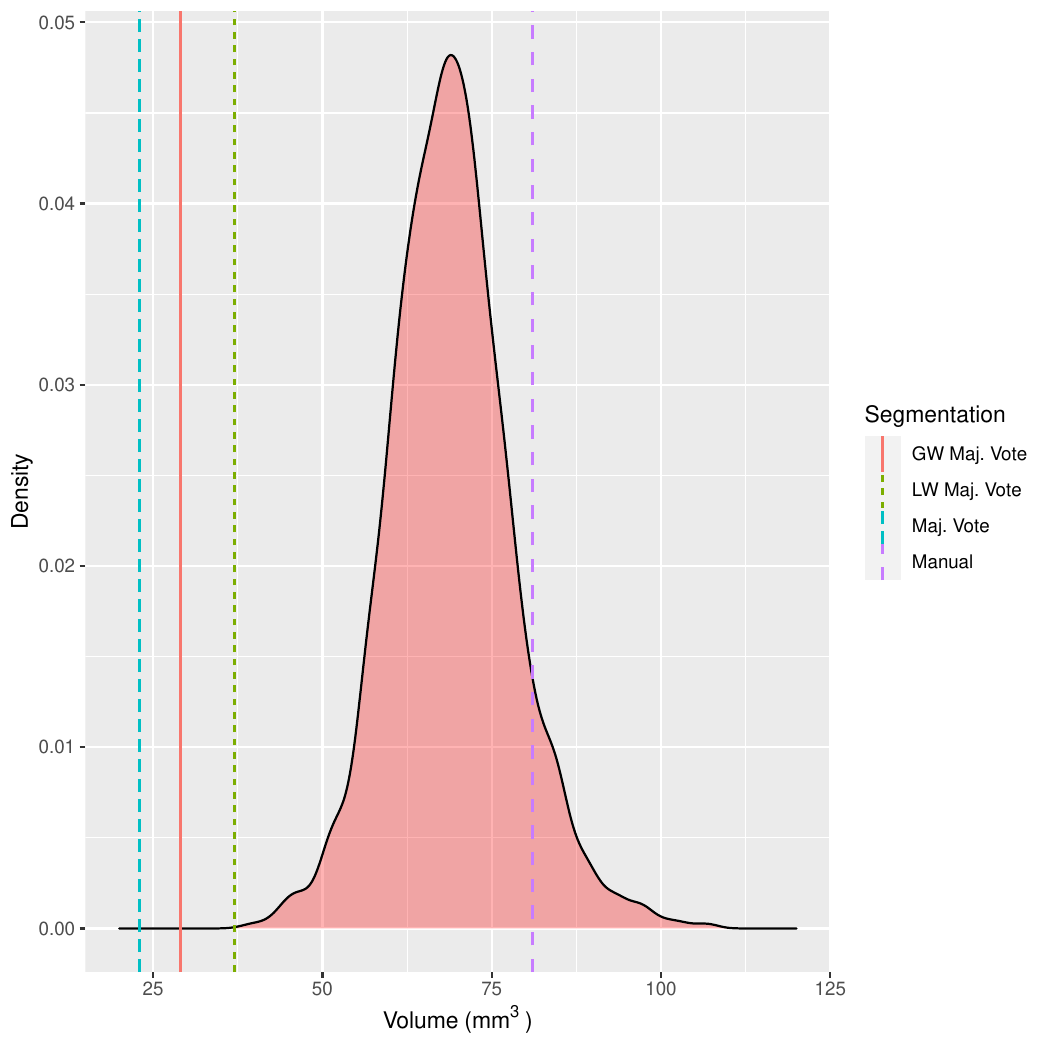}
    \caption{Left four panels: Binary segmentation estimates and Dice coefficents ($D$) obtained via thresholded posterior probability map, simple majority voting, globally weighted (GW) majority voting, and locally weighted (LW) majority voting in the simulated label fusion example. Right panel: Marginal posterior volume density produced from the Bayesian label fusion along with the comparative volume estimates.}
    \label{fig:SigBinMaps}
\end{figure}

Most existing label fusion procedures result in a binary map corresponding to inclusion/exclusion of voxels in a structure of interest. Thresholding the posterior probability maps at 0.5 also yields such a binary map, displayed in Figure \ref{fig:SigBinMaps}. For comparison, the figure also displays the segmentations that result from simple majority voting, globally-weighted majority voting \citep{ArtaEtAl08} in which each atlas is weighted by the inverse of its average squared intensity difference, and locally-weighted majority voting \citep{ArtaEtAl09} in which the contribution of each atlas at each pixel is weighted by the inverse squared intensity difference from the target. We can see the superior recovery of the target under our approach versus different versions of majority voting. The figure also gives the Dice similarity coefficients quantifying similarity between each segmentation and the target, where the target is dichotomized to zero / non-zero pixels. Let $d_v$ denote the indicator of inclusion of pixel $v$ in the estimated image. The Dice coefficient, calculated as $D := 2\sum_{v=1}^Vd_vT_v/(2\sum_{v=1}^Vd_vT_v + \sum_{v=1}^Vd_v(1-T_v) + \sum_{v=1}^V(1-d_v)T_v)$, is often used as a measure of image similarity (though it may not be the best measure when the structure of interest is small relative to the image size \citep{TahaHanbury15}). Dice values close to one indicate strong agreement between two images. Despite the fact that all of the poor segmentations were downweighted relative to the good one, they still outweigh the quality atlas in each voting procedure, resulting in poor segmentation. Thresholding the posterior probability map has resulted in $110\%$ improvement in similarity over even the most favorable version of majority voting. 

Often, a researcher is uninterested in a binary segmentation for its own sake, but rather as a means to an end; e.g., volume estimation. With a binary segmentation, the only way to estimate the volume is by summing the binary indicators over the image. In contrast, our approach facilitates construction of a distribution of plausible volumes. Let $T^{(k)}_v$ denote the realization of $T_v$ on iteration $k$ of the MCMC output. Recognizing that $N^{-1}\sum_{k=1}^N\sum_{v=1}^V T_v^{(k)} \approx \sum_{v=1}^V E(T_v \mid \B{Y}) = \sum_{v=1}^V E[P(T_v = 1 \mid \B{\omega}, \B{Y})]$, we estimate the structure volume on iteration $k$ with $\widehat{M}^{(k)} := \sum_{v=1}^V \widehat{P}(T_v = 1 \mid \B{\omega}^{(k)}, \B{Y}), ~k= 1, \ldots, N$. The right panel in Figure \ref{fig:SigBinMaps} displays the estimated marginal distribution of volume, along with the true volume and the volume estimates obtained from the three voting procedures. The true volume is well within the high probability region of the distribution. We have no way of formally quantifying the uncertainty associated with the voting estimates.

These simulation results demonstrate possible improvements over simple and weighted majority voting through our proposed Bayesian label fusion model. This is possible even when strongly corrupted atlases are used as inputs to the label fusion, since covariate and prior information can protect against otherwise poor segmentaitons. By allowing each atlas' reliability to vary throughout an image, we account for the fact that no atlas is uniformly more reliable than another throughout the image.  Even in the case of quality atlases (e.g., when healthy brains are used to segment healthy brains with a good registration algorithm), we see that including a covariate can improve the segmentation versus no predictors at all. 

\section{Application to Hippocampus Segmentation}
\label{sec:Application}

The Alzheimer's Disease Neuroimaging Initiative (ADNI; \url{http://www.adni.loni.usc.edu}) is a multi-center, longitudinal study with the goals of better understanding Alzheimer's disease (AD) and developing effective biomarkers. The data are publicly available via applying for access, subject to approval by the ADNI Data and Publications Committee. \color{black} Using demographic information available from the ADNI, we created an age- and sex-matched, case-control sample of six AD and six healthy control subjects. Manual segmentations of the left and right hippocampus were obtained for each subject from the Harmonized Protocol For Hippocampal Volumetry \citep{Boc+etal:15}.  The corresponding $T_1$ images were downloaded from the LONI IDA website. For each image, we applied N4 inhomogeneity correction \citep{Tus+etal:10} and Multi-atlas Skull Stripping \citep{Dos+etal:13}. For each target $T_1$ image we consider, the remaining $T_1$ atlases were non-linearly registered to the target image using \verb|SyN| deformable registration with mutual information cost and Welch windowed sinc interpolation \citep{Ava+etal:11} in \verb|R|. The concomitant transformations were applied to each manual segmentation with nearest-neighbor interpolation to obtain the atlases for each target. The $T_1$ intensity values were normalized across images using white stripe \citep{ShinoharaEtAl14}. 


We focus on using known control subjects' brain images as atlases for segmenting the brain of an individual that has been diagnosed with AD. The heterogeneity between the structure of the diseased brain and the healthy controls makes the segmentation task more challenging than in the conventional case. We demonstrate the utility of tissue class information in addition to atlas-target image agreement. After considering the single-brain case, we summarize the performance of our proposed approach over all six AD brain images.

\subsection{Segmenting the Hippocampus of a Single Diseased Brain}
\begin{figure}[tb]
    \centering
    \includegraphics[scale= 0.3, trim= {0 25cm 0 0},clip]{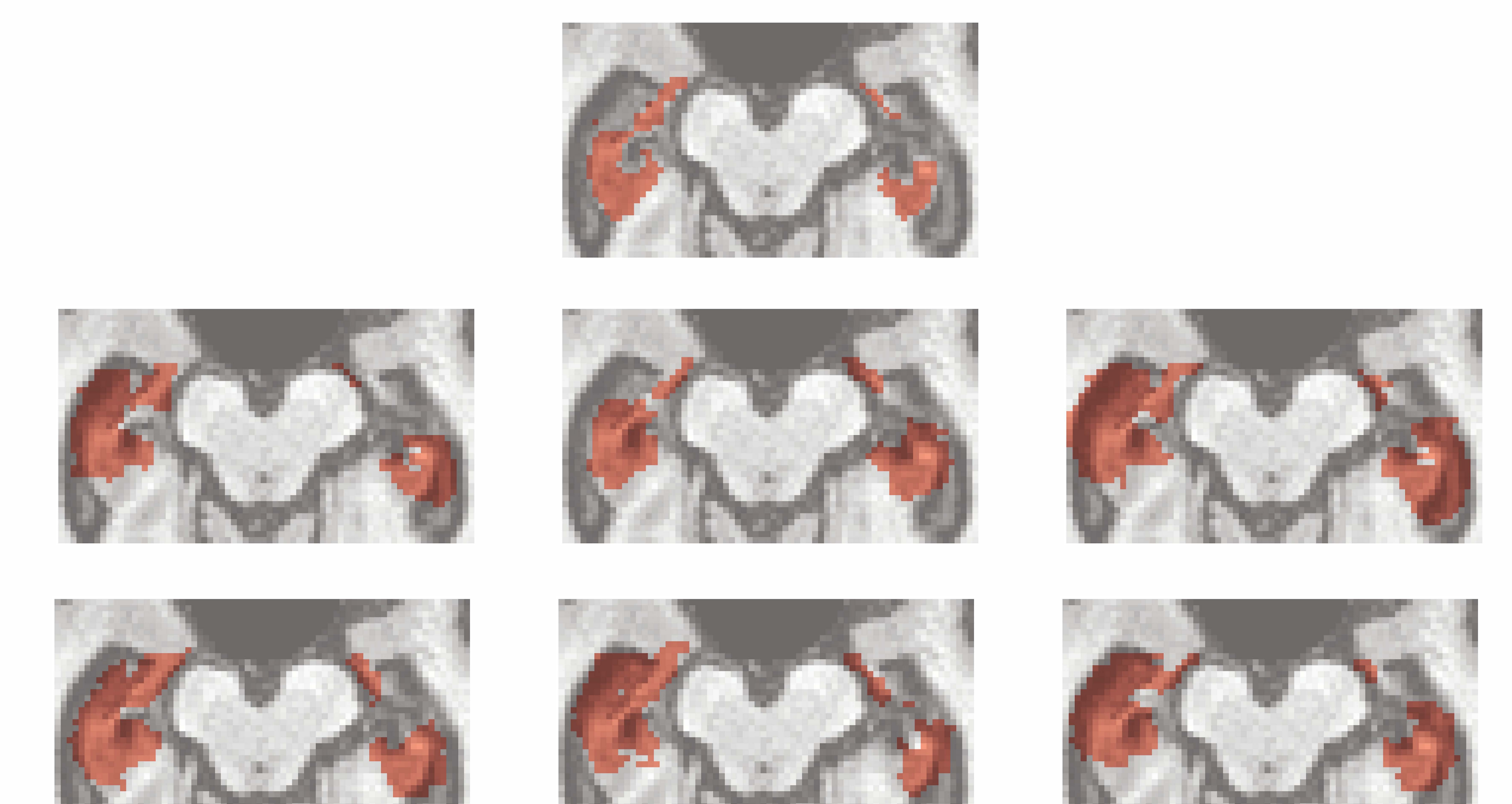}
    \caption{Axial view of one slice of the manually segmented hippocampus (top) and six atlas segmentations (bottom two rows) for the label fusion example. The target brain has been diagnosed with AD; the atlases come from healthy controls.}
    \label{fig:atlas_segmentations}
\end{figure}

We take as our target the three-dimensional \color{black}  brain image of a 78 year old male diagnosed with AD.  The six controls are registered to the target brain. 
To reduce the size for the sake of computation without losing information about the plausible location and volume of the hippocampus of interest, the images are cut to three-dimensional rectangles of the same size, where the rectangle is the smallest box containing the largest segmentation. This results in each image having dimension $70 \times 39 \times 27$ so that $V = 73,710$ and $R = 6$ in \eqref{eqn:datDist}. One representative slice from each of the resulting three-dimensional atlas segmentations is displayed in Figure \ref{fig:atlas_segmentations}, along with the same slice from each of the manual segmentations. \color{black}  In this figure we observe a systematic tendency for each atlas to considerably over-segment the hippocampus when compared to the manual segmentation. This is due in part to the fact that each atlas is based on a brain that has not been diagnosed with AD. It is well established that hippocampal atrophy is more severe in AD patients compared to a healthy population. Thus, the hippocampi from the reference brains tend to be larger than than that in the target image; a systematic discrepancy that persists even after registering the labels to target image space. These discrepancies between the atlases and the target create challenges for label fusion approaches.

For covariate information, we take the weighted sum of the SDL maps for each atlas (rescaled to the unit interval), where each atlas is weighted voxelwise by its $T_1$ intensity similarity with the target $T_1$ image.  The use of atlas-target image agreement is common among modern label fusion algorithms and has proven to be useful, so we use it here. To further guide the segmentation, however, we recognize that the hippocampus is a structure consisting only of gray matter. Thus we use a pre-computed tissue class segmentation to compute voxelwise gray matter indicators for inclusion in our regression model. The tissue class segmentation is obtained via the \texttt{ATROPOS} algorithm \citep{Atropos11}. The SDL map and tissue class segmentation, along with the manually-segmented target image, are displayed in Supplementary Figure 4. We include an interaction term to downweight the effect of the presence of gray matter when a voxel is thought to be far away from the hippocampus. To elicit the prior on the regression coefficients via CMP, we use pseudodata to impose our prior knowledge that (i) gray matter and a small SDL has a very high likelihood of truly belonging to the hippocampus, (ii) a large SDL and not gray matter is very likely to be external to the hippocampus, (iii) gray matter with a large SDL is somewhat likely external to the hippocampus, and (iv) a small SDL but segmented as not belonging to gray matter has a marginal probability of being part of the hippocampus, due to possibly misclassified tissue type. (See Supplmentary Material for additional details.) The priors on the sensitivity and specificity processes are specified as they were in Section \ref{sec:Simulation}. 

After initialization, we run 150,000 \color{black}     iterations of a single Monte Carlo Markov chain, thinning to every $50^{\text{th}}$ draw. The last 1,500 \color{black}  realizations are retained as an approximate sample from the posterior distribution. We examine trace plots and ergodic averages of the $\B{\delta}$ coefficients as a convergence diagnostic. Supplementary Figure 5 displays these plots along with the Geweke statistics and lag 1 autocorrelations, all of which suggest convergence. The MCMC routine is coded in \texttt{R} and \texttt{C++} via the \texttt{Rcpp} package \citep{RcppPack} and executed on a Dell Precision T3620 desktop PC running Windows 10 with an Intel Xeon 4.1 GHz CPU and 64GB RAM. It takes approximately 17 hours to run.\color{black}

\begin{figure}[tb]
    \centering
    \includegraphics[scale=.25, trim= 0.5in 1.25in 0.5in 1.25in]{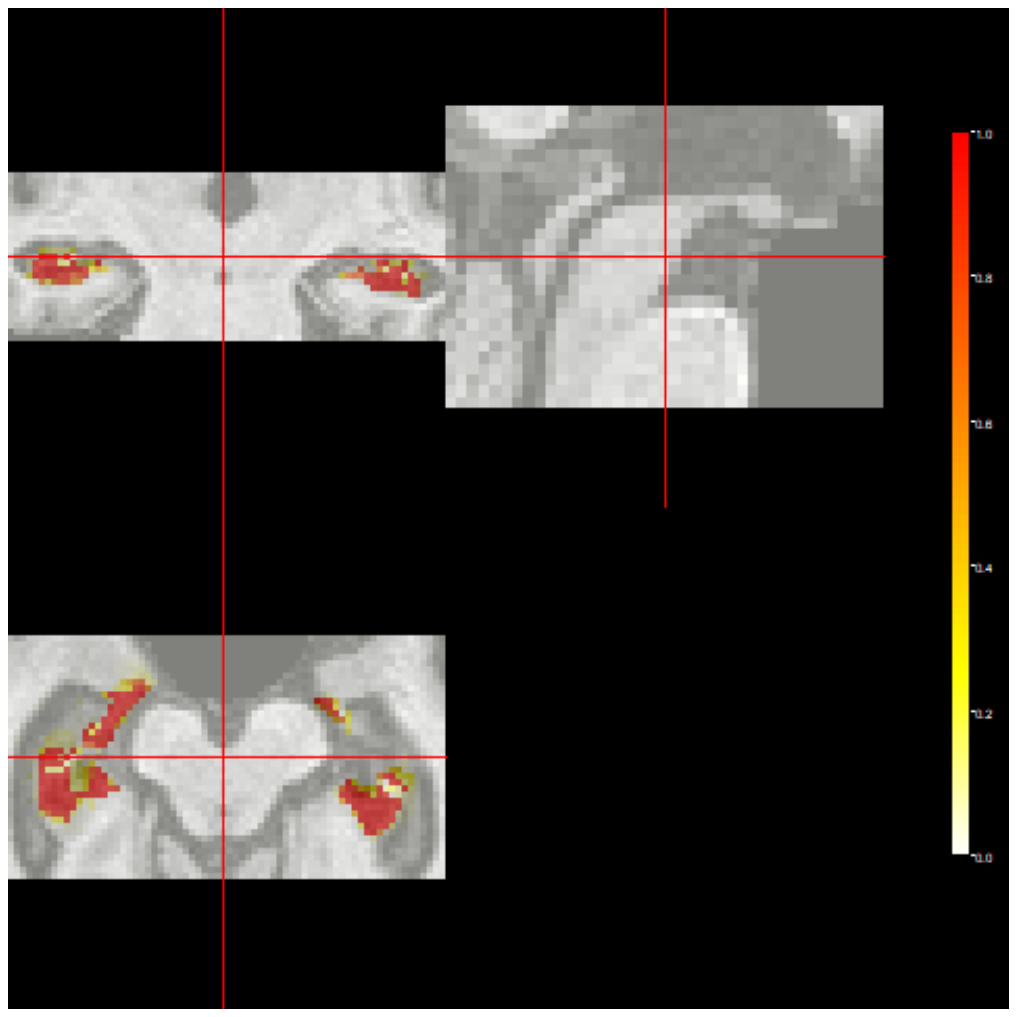} \hspace{60pt}
    \includegraphics[scale=.25, trim= 0.5in 1.25in 0.5in 1.25in]{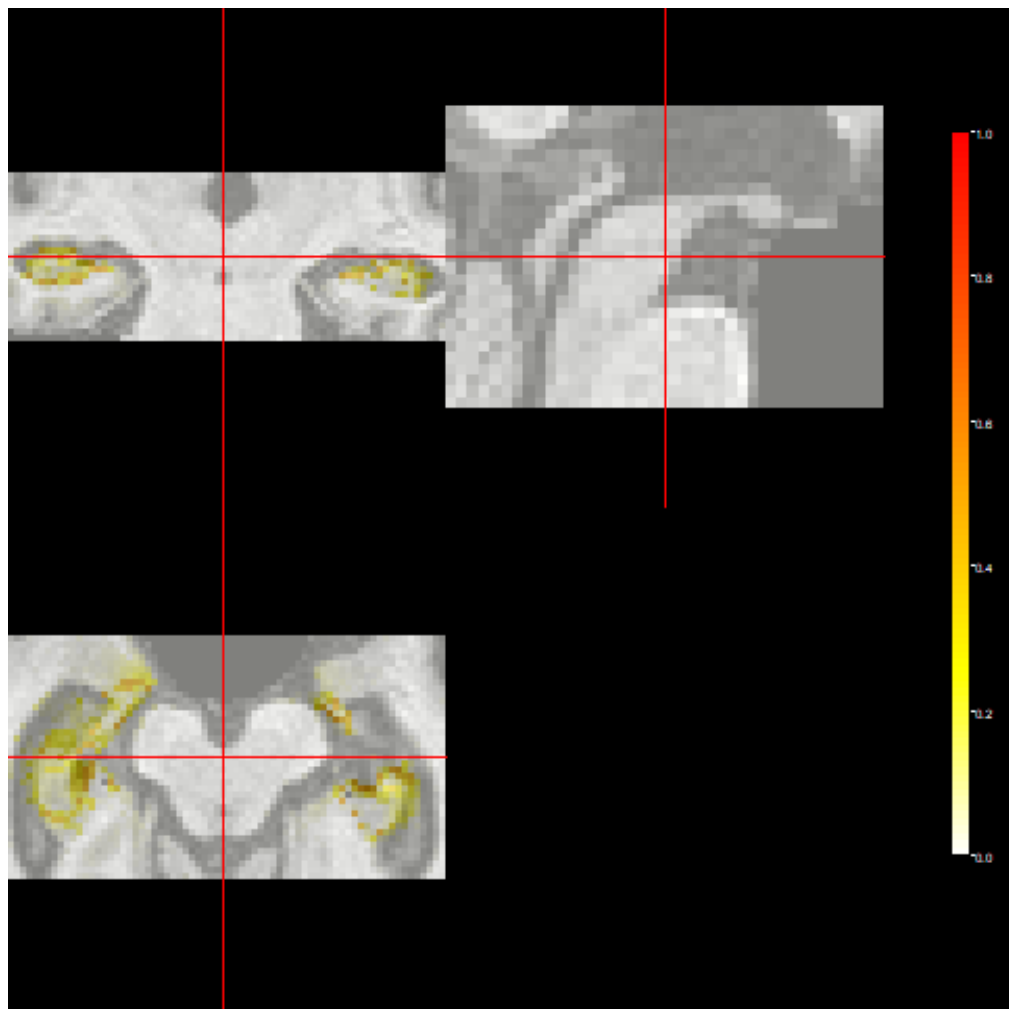}
    \caption{Posterior inclusion probability map (left) and pointwise standard deviations (right) for segmenting the hippocampus of an AD  patient. In each 3D orthographic image the views are coronal (top left), sagittal (top right), and axial (bottom left). Note that each image displays only the subset of the full brain image that was used for the label fusion --- the smallest rectangle containing the largest segmentation. The coronal view displays no structure since the slice shown is in between the left and right hippocampus.}
    \label{fig:ADNIppm}
\end{figure}

Figure \ref{fig:ADNIppm} displays the posterior probability map for hippocampus inclusion, along with the point-wise standard deviations. The highest probabilities of inclusion are near the centers of the two regions. There is considerable uncertainty about the edges. If a researcher were only interested in the hippocampal volume for the patient, these probabilities could be used directly to estimate the volume, as discussed below.

\begin{figure}[tb]
    \centering
    \includegraphics[trim={0 10cm 0 0},clip,scale=.275]{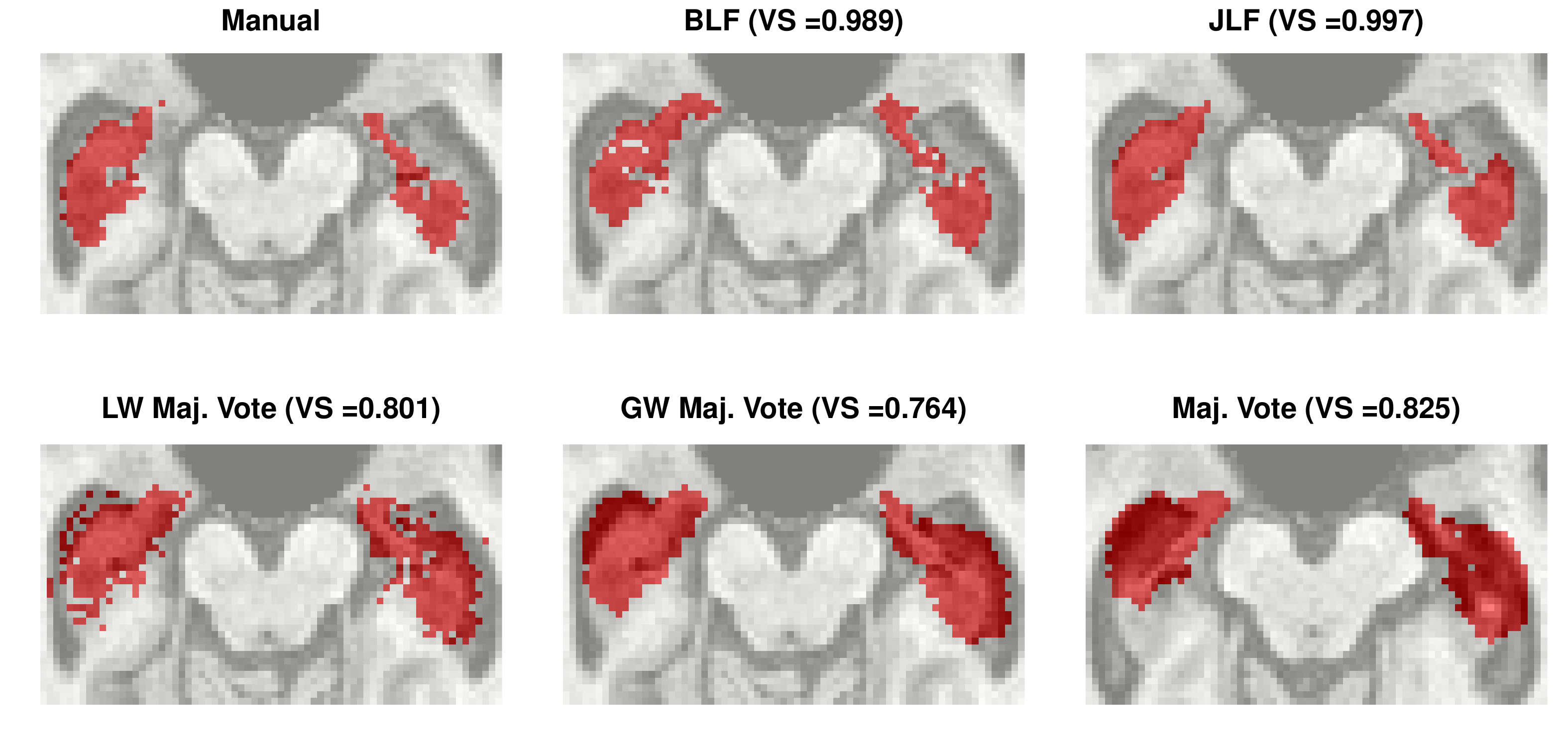}
    \caption{Representative slice from 3D segmentations in the hippocampus segmentation. Manual (top left) and automatic segmentations produced by the proposed Bayesian label fusion (BLF, top middle) model, joint label fusion (JLF, top right), locally-weighted (LW, bottom left) majority voting, globally-weighted (GW, bottom middle) majority voting, and simple majority voting (bottom right). Displayed above each segmentation is the volume similarity (VS) between the automatic and manual segmentations.}
    \label{fig:0592MASegs}
\end{figure}

One can threshold the posterior probabilities to obtain a binary inclusion map. After thresholding, we compare the resulting segmentation to those obtained by simple majority voting, globally-weighted majority voting, locally-weighted majority voting, and JLF. The global weighting is inversely proportional to each atlas' average $T_1$ intensity difference from the target. Local weighting is done similarly using voxel-specific intensity differences. Figure \ref{fig:0592MASegs} displays one slice of the manual segmentation along with that which is obtained by thresholding the posterior inclusion probabilities at 0.5. The hippocampus is a relatively small structure compared to the full three-dimensional image (as can be seen in, e.g., Supplementary Figure 4). In this case, \cite{TahaHanbury15} argue that the Dice coefficient defined in Section \ref{sec:Simulation} is not the best measure for evaluating a segmentation. However, the {\em volume} of the hippocampus is important for volumetry in the study of AD progression. Thus, we follow the suggestion of \cite{TahaHanbury15} and use the volume similarity as an evaluative metric, defined as $VS = 1 - |FN - FP|/(2TP + FP + FN)$, where $FN, FP,$ and $TP$ denote false negatives, false positives, and true positives, respectively. Bayesian label fusion attains $VS = 0.989$. This is competitive with JLF, the current state-of-the-art ($VS =0.997$), and superior to both simple and weighted voting procedures. (VS values are displayed in Figure \ref{fig:0592MASegs}.) 

Unlike the other automatic segmentations, the fully Bayesian label fusion produces measures of uncertainty in the form of a posterior probability distribution that can be used to derive marginal distributions of any quantity of interest, as we discuss below. Further, our proposed model also has the ability to estimate the spatially-varying reliability parameters for each atlas. Supplementary Figure 6 displays the sensitivity and specificity maps for one selected atlas. We see smooth decay of sensitivity between regions of high agreement and low agreement with the manual segmentations. Since the sensitivity is conditional on the voxel truly being part of the structure ($T_v = 1$), the model is only capable of estimating sensitivity in areas where it estimates a high probability of the voxel truly being part of the structure. The fields return to their prior means as they move away from the structure estimate. Similarly, the specificity values are estimated to be very high away from the targeted structure, where all atlases agree on the voxels being excluded.

\begin{figure}[tb]
    \centering
    \includegraphics[scale= 0.25, trim= 0.5in 1.25in 0.5in 1.25in]{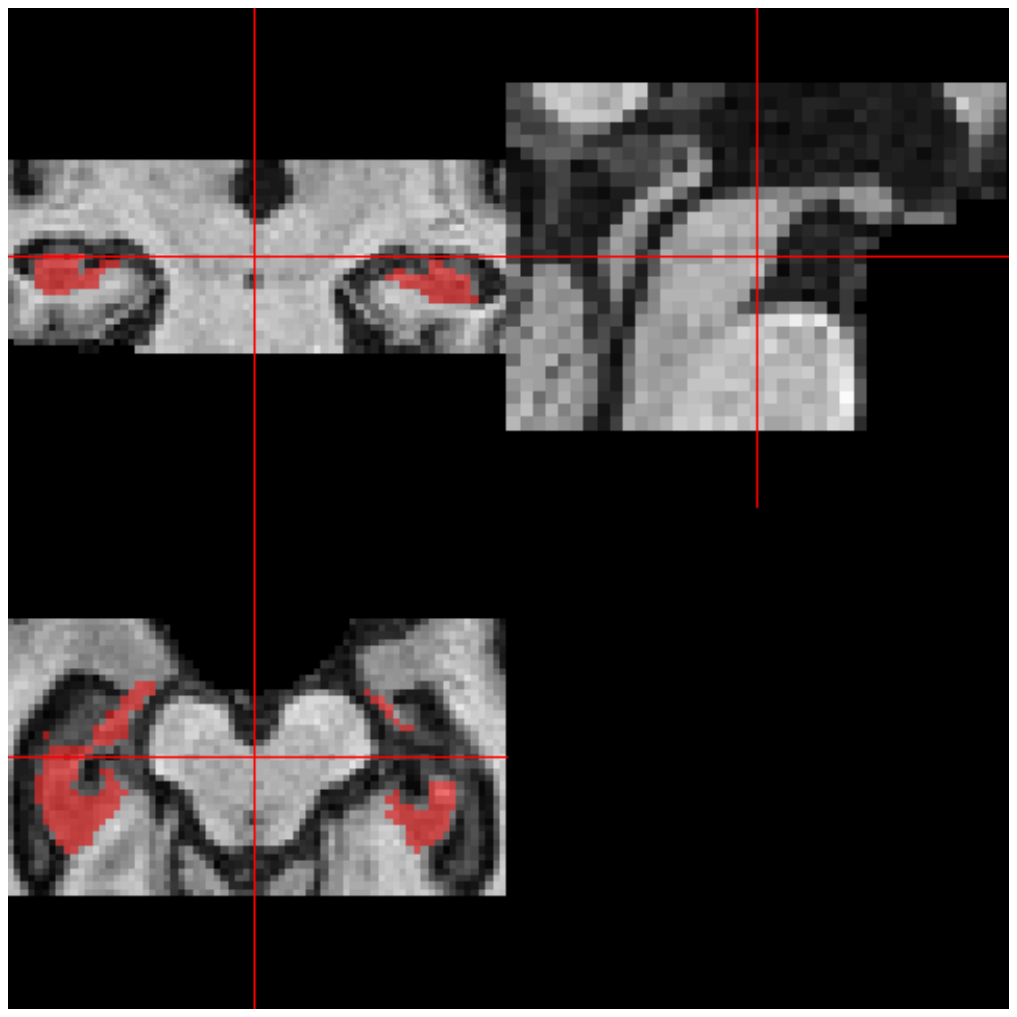}\hspace{8pt}
    \includegraphics[scale= 0.25, trim= 0.5in 1.25in 0.5in 1.25in]{PostMeanOrtho.pdf}\hspace{8pt}
    \includegraphics[scale= 0.25, trim= 0.5in 1.25in 0.5in 1.25in]{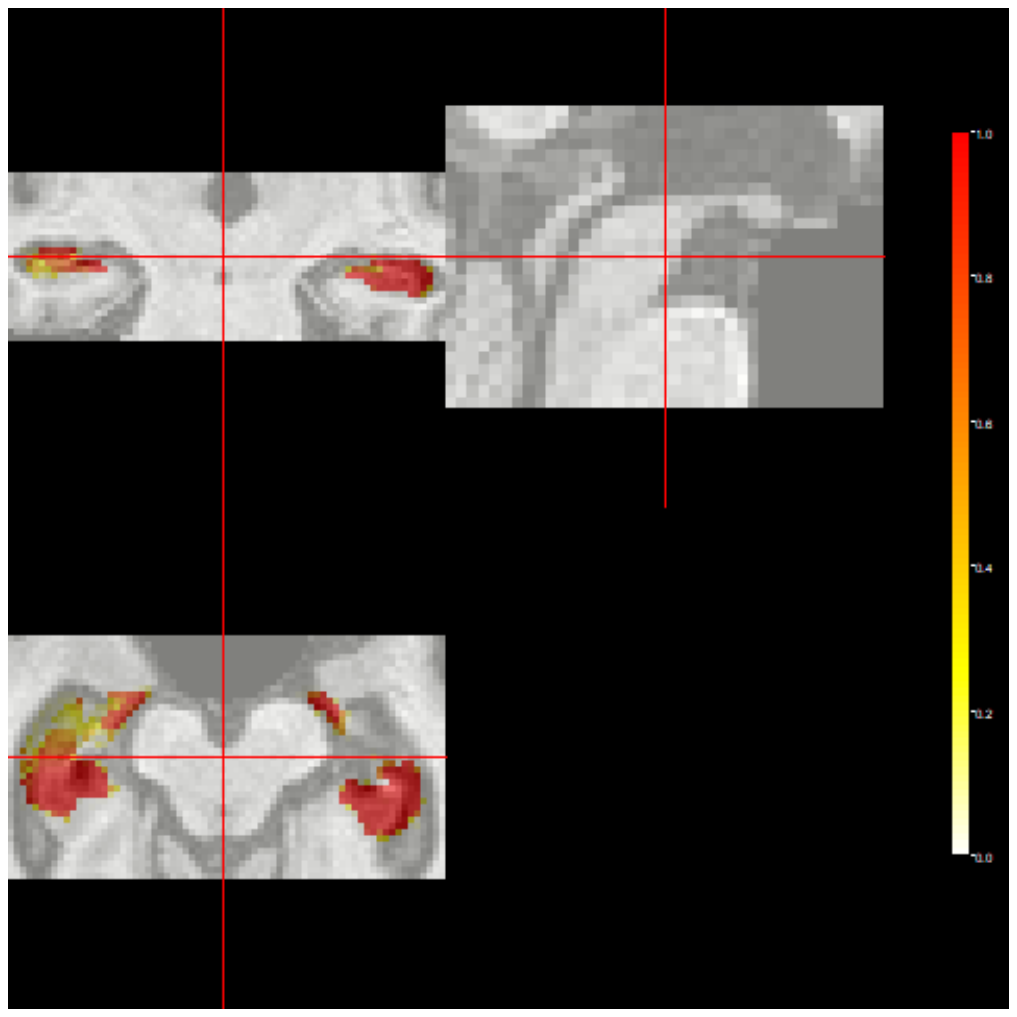}
    \caption{Manual segmentation (left) along with posterior probability maps obtained with (center) and without (right) using a gray matter tissue class segmentation as a covariate in the Bayesian label fusion model. Note that each image displays only the subset of the full brain image that was used for the label fusion --- the smallest rectangle containing the largest segmentation. The coronal view displays no structure since the slice shown is in between the left and right hippocampus.}
    \label{fig:0592SDLsAndGM}
\end{figure}

\color{black}
As already mentioned, only healthy brains are used as atlases. The systematic differences are not completely captured by atlas-target image dissimilarities. Despite accounting for image dissimilarity, there is a tendency for the established methods to over-segment the hippocampus. The Bayesian label fusion model facilitates explicit incorporation of the estimated gray matter pattern as a predictor. Our prior specification allows for the possibility that the tissue classes are incorrectly assigned in some places, but are mostly reliable. The effect of the gray matter segmentation as auxiliary information can be clearly seen by comparing it even with our own model in which this information is ignored but the model is otherwise identical. Figure \ref{fig:0592SDLsAndGM} displays the posterior inclusion probabilities obtained without using the gray matter information, along with the results already presented for reference. Using only the intensity-similarity-weighted distance labels yields a Bayesian regression analogue to the other approaches that only weight by intensity similarity. The additional tissue class information is able to prevent oversegmentation of the diseased structure. 

\begin{figure}[tb]
    \centering
    \includegraphics[scale=.35]{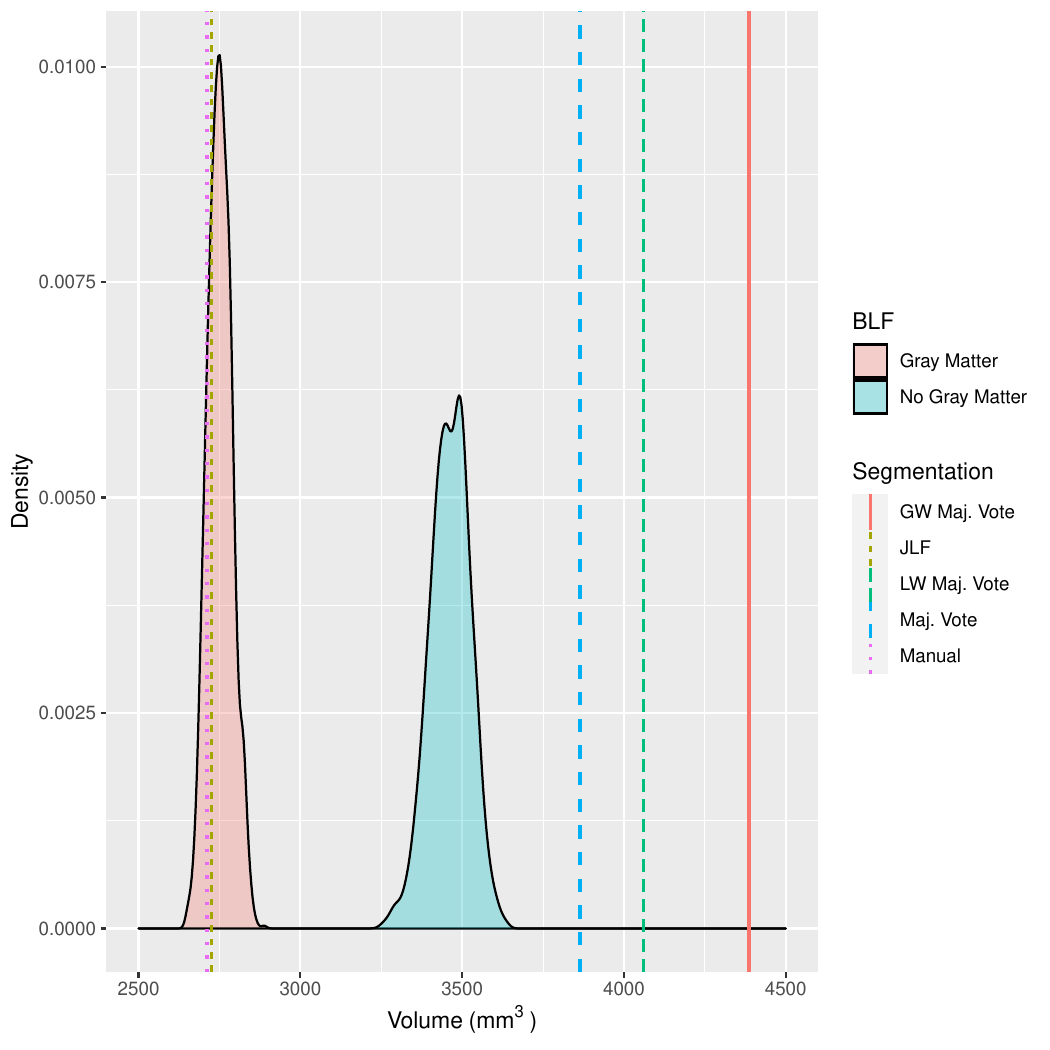}
    \caption{Estimated marginal posterior distribution of  hippocampus volume produced by the proposed Bayesian label fusion model with and without gray matter information. Also plotted are the manual, joint label fusion (JLF), locally-weighted (LW), globally-weighted (GW), and simple majority voting volume estimates.}
    \label{fig:0592VolDist}
\end{figure}

In practice, an anatomical structure is segmented to obtain important information such as its volume or average image intensity within the structure. In our case, segmenting the hippocampus is a step toward estimating its volume. If we only obtain a binary map, then the only way to estimate the volume is by summing the indicators. Doing so ignores many sources of uncertainty, including image pre-processing, registration error, biological variation, and rater variability. Monte Carlo sampling also facilitates estimation of a distribution of plausible volumes through $\hat{M}^{(k)}$, defined in Section \ref{sec:Simulation}. Figure \ref{fig:0592VolDist} displays the marginal volume densities for the diseased brain of interest, both with and without gray matter included in the Bayesian label fusion model. Vertical lines indicate the manually-segmented volume and the volume estimates from the other procedures considered.  The benefit of including the gray matter information is again evident with the improved agreement of the volume distribution with the manual segmentation. In this case, though, we see that even without the gray matter information the Bayesian model outperforms the three majority voting procedures.\color{black}

\subsection{Aggregated Results}\label{sec:aggRes}
\begin{figure}[tbp]
    \centering
    \includegraphics[scale= 0.4, clip= TRUE]{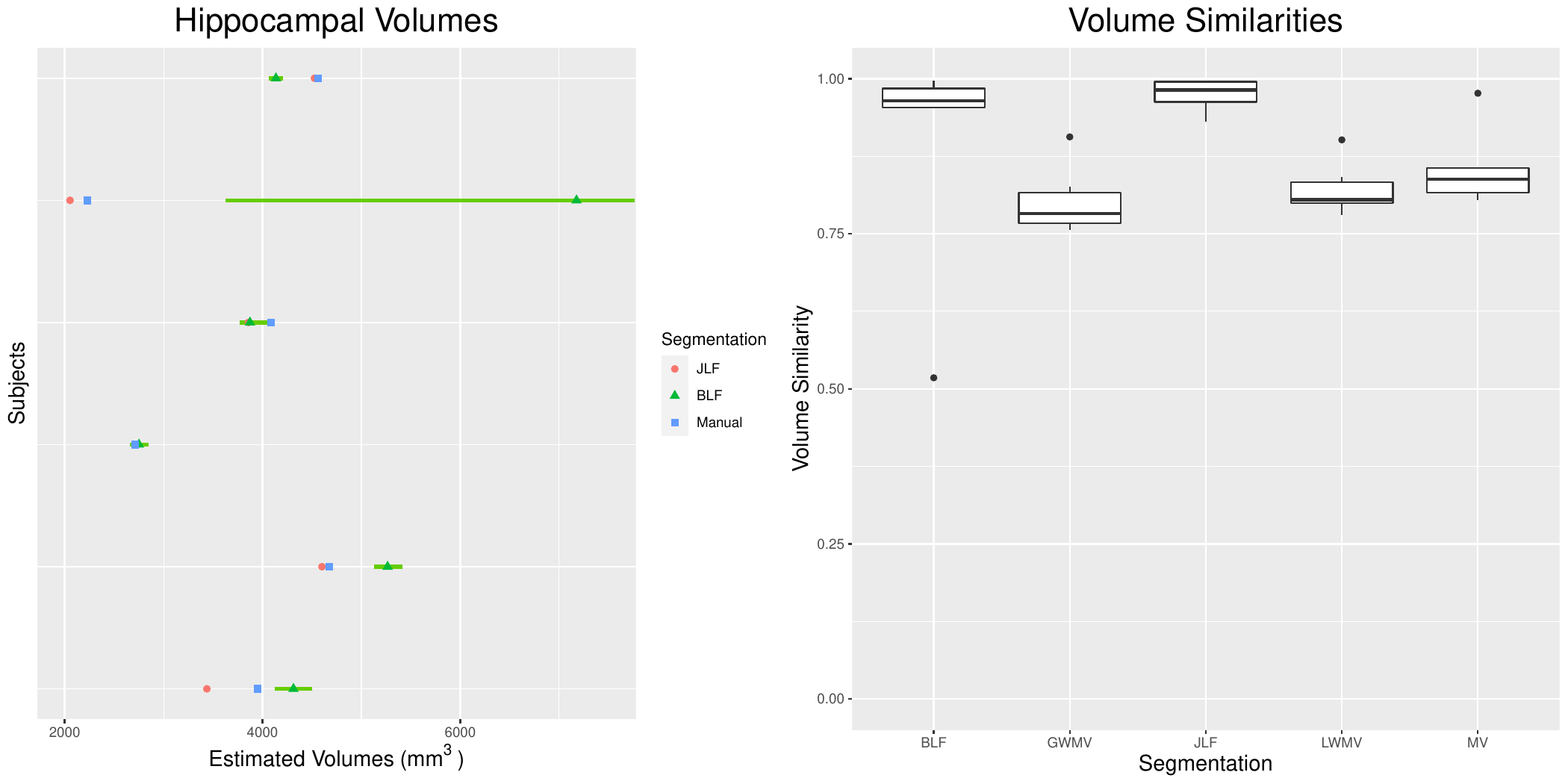}\\
    \caption{Summary of volume estimates (left) and volume similarities (right) with respect to the manual segmentations for the proposed Bayesian label fusion (BLF), joint label fusion (JLF), locally-weighted majority voting (LWMV), globally-weighted majority voting (GWMV), and simple majority voting (MV) when using control cases as atlases for the AD patients. The bars around the BLF estimates in the left panel depict approximate 99\% credible intervals. For simplicity, only the JLF and BLF volume estimates are displayed in the left plot.}\label{fig:aggResults}
\end{figure}

Here we present the  results of using the control brain images as atlases for each of the six AD patients in our dataset. Figure \ref{fig:aggResults} summarizes the volume estimates and volume similarities between the automatic and manual segmentations. In terms of volume similarity, our approach and JLF separate themselves from the three voting approaches. In addition to being competitive with the current state of the art, our approach is the only one that meaningfully produces measures of associated uncertainty, depicted in this case as 99\% posterior credible intervals. This uncertainty is particularly evident in the second subject from the top depicted in the left panel of Figure \ref{fig:aggResults}. The point estimate (posterior mean) is quite far from the manually segmented volume relative to differences in the other brain images. However, the posterior distribution indicates a large amount of uncertainty about this particular estimate. Indeed, the MCMC trace plot of volume estimates for this subject (labeled 1263), displayed in Supplementary Figure 6, suggest that the marginal posterior distribution of volume is bi-modal. This is a feature of the posterior distribution that an optimization routine would likely miss. \color{black} Lastly, we remark that even a trained expert will produce slightly different manual segmentations of the brain on two different occasions, so there is uncertainty associated with each manual segmentation itself. This is not quantified here. Thus it is impossible to assess any `significant' difference between volume estimates from our proposed approach and a manually estimated volume.

These results demonstrate the feasibility of our approach on a full three-dimensional segmentation \color{black} as well as additional useful information that would be unavailable otherwise. Posterior probability maps can be used to summarize where the hippocampus is likely to be. A simple thresholding rule yields image similarities that are competitive with the state-of-the-art when using healthy brains as atlases for diseased cases. However, thresholding is unnecessary when the goal is to estimate the volume of the hippocampus, as shown by the availability of a posterior volume distribution. Incorporating tissue class information into a generative Bayesian model for segmentation results in more faithful estimates of the hippocampal volume along with meaningful measures of uncertainty.

\color{black}
\section{Discussion}
\label{sec:Discussion}

We have taken a hierarchical Bayesian approach to label fusion for anatomical segmentation. We proposed a generative regression model to incorporate additional information to guide label fusion, information that is not typically used in existing procedures. In particular, we demonstrated considerable improvements in healthy-on-diseased label fusion that can be achieved by considering voxelwise tissue classifications in addition to image similarity. The resulting segmentations from our approach are competitive with the current state-of-the-art and were shown to outperform other popular voting-based procedures. Further, however, we were able to fully account for all sources of uncertainty by accessing the posterior distribution via MCMC. With posterior probability maps and their associated distributions, a researcher can threshold to obtain a binary segementation. The approach also facilitates estimating a distribution of plausible volumes. Access to such information can be important for assessing statistically significant differences between subjects or within a subject over time, since point estimates alone can be misleading \citep{FoxNicholls01, GreenlawEtAl17}. 

In the aggregated analysis in Section \ref{sec:Application}, our assumed regression model and prior weighting was the same across each brain image. However, the full three-dimensional images are different from each other both in terms of image size (i.e., sizes of the rectangles containing the largest segmentation) and voxel size. (Some of the images have voxels of dimension $1.2mm \times 1.25mm \times 1.25mm$ and others have voxels of dimension $1.2mm \times 0.9375mm \times 0.9375mm$.) The former issue affects appropriate prior weighting in the CMP prior, as our experience is that the appropriate number of pseudo-observations is relative to the image size. The latter issue suggests that a more flexible, anisotropic spatial model for the reliability fields might be more appropriate than the CAR model we used in this work for wider applicability. Lastly, while we found gray matter indicators to be useful for the case considered here, other atlas-independent information and/or prior elicitation could produce stronger results in other situations with different populations of interest. We defer these issues to future investigation. \color{black}

We demonstrated in this work the feasibility of fully Bayesian modeling via MCMC on three dimensional brain images. This is noteworthy since many MCMC-based Bayesian approaches in the neuroimaging literature have been demonstrated on single two-dimensional slices. \color{black} Regardless, for even larger scale studies than that considered in this work, there exists in the literature a variety of techniques for MCMC in high dimensional spaces, including when the model involves GMRFs \citep[e.g.,][]{SidenEtAl17}. Additional options are available if one is willing to move away from GMRFs. For instance, \citet{BezenerEtAl18} use pre-defined regions to aggregate voxels and reduce the dimension of the spatial field underlying fMRI data while maintaining spatial dependence. \citet{ZhaoEtAl18} mitigate the computational burden with a multi-resolution MCMC approach that successively refines resolutions in interesting areas of a brain image. \citet{HetatonEtAl18} review many large scale Gaussian process techniques that have been proposed recently. Also available are posterior approximation methods such as variational Bayes, though such approximations can yield inferior results to those produced by MCMC \citep{SidenEtAl17, TengEtAl16}. Concerning the extension from binary segmentation to whole-brain parcellation, an obvious modification to our proposed approach is to replace the binomial response with multinomial. This raises new challenges, not the least of which is computation. Given the aformentioned MCMC advances, though, we are optimistic that such a path is within reach.





 \section{Acknowledgements}
 Data collection and sharing for this project was funded by the Alzheimer's Disease Neuroimaging Initiative (ADNI) (National Institutes of Health Grant U01 AG024904) and DOD ADNI (Department of Defense award number W81XWH-12-2-0012). ADNI is funded by the National Institute on Aging, the National Institute of Biomedical Imaging and Bioengineering, and generous private sector contributions. This material is based upon work partially supported by the National Science Foundation (NSF) under Grant DMS-1127914 to the Statistical and Applied Mathematical Sciences Institute. DAB is partially supported by NSF Grants CMMI-1563435 and OIA-1826715. CSM is partially supported by National Institutes of Health Grant AI121351, NSF Grant OIA-1826715, and the Department of Defense's Office of Naval Research Grant N00014-19-1-2295. RTS is partially supported by NIH Grants R01NS085211, R21NS093349, R01MH112847, R01NS060910, and R01EB017255. The content is solely the responsibility of the authors and does not necessarily represent the official views of the funding agencies. The authors thank the Editor, an Associate Editor, anonymous referees, and the clinical neuroimaging research group in the Department of Biostatistics at Johns Hopkins University, particularly John Muschelli for his help with the image registrations.

\bibliographystyle{apa}
\bibliography{citesR3}

\end{document}